\documentclass[fleqn,usenatbib]{mnras}

\usepackage{newtxtext,newtxmath}
\usepackage{setspace}

\usepackage[T1]{fontenc}

\DeclareRobustCommand{\VAN}[3]{#2}
\let\VANthebibliography\thebibliography
\def\thebibliography{\DeclareRobustCommand{\VAN}[3]{##3}\VANthebibliography}

\usepackage[utf8]{inputenc}
\usepackage[english]{babel}
\usepackage{listings}
\usepackage{longtable}

\usepackage{graphicx}	% Including figure files
\usepackage{bm}

\usepackage[usenames,dvipsnames]{xcolor}
\usepackage{ulem}
\usepackage{hyperref}
\usepackage{subfloat}

\newcommand{\kms}{\mbox{km\,s$^{-1}$}}	% km s^-1
\newcommand{\kmskpc}{\mbox{km\,s$^{-1}$\,kpc$^{-1}$}}	% km s^-1 kpc^-1

\newcommand{\Vh}{V_{\rm h}}

\newcommand{\LCDM}{$\Lambda$CDM}
\newcommand{\DMW}{D_\mathrm{MW}}
\newcommand{\Msun}{\mbox{$\mathrm{M}_\odot$}}
\newcommand{\MLMC}{\mbox{$\mathrm{M}_\mathrm{LMC}$}}

\title[The LMC impact on the MW satellites]{The LMC impact on the kinematics of the Milky Way satellites: clues from the running solar apex}
\author[Makarov et al.]{
Dmitry Makarov$^{1}$,
Sergey Khoperskov$^{2,3}$,
Danila Makarov$^{1}$,
Lidia Makarova$^{1}$,
Noam Libeskind$^{2}$,
\newauthor
Jean-Baptiste Salomon$^{4,2,5}$
\\
$^1$Special Astrophysical Observatory, Russian Academy of Sciences, Nizhnii Arkhyz, 369167 Russia\\
$^2$Leibniz Institut f\"{u}r Astrophysik Potsdam (AIP), An der Sternwarte 16, D-14482, Potsdam, Germany\\
$^3$GEPI, Observatoire de Paris, PSL Universit{\'e}, CNRS,  5 Place Jules Janssen, 92190 Meudon, France\\
$^4$Racah Institute of Physics, Hebrew University, Jerusalem 91904, Israel\\
$^5$Institut UTINAM, CNRS UMR6213, Univ.\ Bourgogne Franche-Comté, OSU THETA, Observatoire de Besançon, BP 1615, 25010 Besançon Cédex, France
}

\date{Accepted XXX. Received YYY; in original form ZZZ}

\pubyear{2022}

\begin{document}
\label{firstpage}
\pagerange{\pageref{firstpage}--\pageref{lastpage}}
\maketitle

\begin{abstract}
Dwarf galaxies provide a unique opportunity for studying the evolution of the Milky Way~(MW) and the Local Group as a whole.
Analysing the running solar apex based on the kinematics of the MW satellites, we discovered an unexpected behaviour of the dipole term of the radial velocity distribution as a function of the galactocentric distance.
The nearby satellites~($<100$~kpc) have a bulk motion with an amplitude of 140--230~\kms{} while the more distant ones show an isotropic distribution of the radial velocities.
Such strong solar apex variations can not be explained by the net rotation of the satellites, as it would require an enormously high rotation rate~($\approx 970$~\kms). 
If we exclude the LMC and its most closely related members from our sample, this does not suppress the bulk motion of the nearby satellites strongly enough.
Nevertheless, we have demonstrated that the observed peculiar kinematics of the MW satellites can be explained by a perturbation caused by the first infall of the LMC. 
First, we `undone' the effect of the perturbation by integrating the orbits of the MW satellites backwards~(forwards) with~(without) massive LMC. 
It appears that the present-day peculiar enhancement of the solar apex in the inner halo is diminished the most in the case of $2\times10^{11}$~\Msun{} LMC. 
Next, in self-consistent high-resolution $N$-body simulations of the MW-LMC interaction, we found that the solar apex shows the observed behaviour only for the halo particles with substantial angular momentum, comparable to that of the MW satellites. 
\end{abstract}


%%%%%%%%%%%%%%%%%%%%%%%%%%%%%%%%%%%%%%%%%%%%%%%%%%

%%%%%%%%%%%%%%%%% BODY OF PAPER %%%%%%%%%%%%%%%%%%

\section{Introduction}

Owing to its proximity, the Local Group is one of the most opportunistic systems of galaxies and, 
thus, it is a testbed for various problems of modern galaxy formation theories. 
Over the past decades various large-scale surveys have led to the discovery and detailed study of several dozen of nearby dwarf galaxies~\citep[see a review by][and references therein]{2019ARA&A..57..375S}.
Thanks to the \textit{Gaia} space mission, proper motions have been measured for most of the nearby satellites~\citep[for instance, see recent works by][]{2020AJ....160..124M,2021ApJ...922...93H,2022ApJ...940..136P}. These data, together with the line-of-sight radial velocities, provide an opportunity to study both the formation and evolution of the Local Group galaxies. 

It is assumed that objects inside the virial regions of groups must be `randomised', and coherent structures cannot persist for a very long time.
However, \citet{1976RGOB..182..241K} and \citet{1976MNRAS.174..695L} noted 
that the Milky Way~(MW) satellites are arranged in a wide belt along a great circle perpendicular to the disc of the Galaxy. \citet{2005A&A...431..517K,2007MNRAS.374.1125M,2012MNRAS.423.1109P} show that the flattened distribution of the MW satellites is $99.5$ per cent inconsistent with an isotropic or prolate distribution of substructures, as might be expected in \LCDM. 
Similar planar structures were found around all nearby massive galaxies: the Andromeda galaxy~\citep{2007MNRAS.374.1125M}, M~81~\citep{2013AJ....146..126C}, Centaurus~A~\citep{2015ApJ...802L..25T}, and NGC~253 \citep{2021A&A...652A..48M}. 
Moreover, observations demonstrate that the plane of satellite galaxies, perpendicular to the MW disc, can be rotationally supported~\citep{2013MNRAS.435.2116P}. 
There are also evidences in favour of a regular rotation of the plane of satellites around the Andromeda galaxy~\citep{2013Natur.493...62I}. 
Therefore, planes of satellites may be quite widespread in the nearby Universe. However, modern galaxy formation simulations are not in favour of the stability of these structures~\citep[see, e.g.,][]{2021NatAs...5.1185P}.

One possible explanation of the flat structures in the Local Group is a natural outcome of the anisotropic infall of satellites along dark matter (DM) filaments of the cosmic web \citep{2004MNRAS.352..376A, 2005MNRAS.363..146L}.  
It is an inherent process of halo formation in \LCDM{} cosmology. 
Indeed, \citet{2007ApJ...668..949B} concluded that the Large and Small Magellanic Clouds~(LMC, SMC) are on their first passage around the MW~\citep[see also][]{2010ApJ...721L..97B, 2012MNRAS.421.2109B, 2013ApJ...764..161K}. 

The present-day estimates of the LMC mass suggest that it is a very massive galaxy~($\propto 10^{11}~\Msun$) affecting the motions of stars in the stellar halo of the MW~\citep{2021Natur.592..534C,2021MNRAS.506.2677E}, shaping properties of surrounding stellar streams~\citep{2018MNRAS.481.3148E,2019MNRAS.487.2685E}, globular clusters~\citep{2020MNRAS.499..804G}, and waking the DM distribution on large scales~\citep{2019ApJ...884...51G,2020ApJ...898....4C}. As a result, the infalling LMC, together with its satellite galaxies escort, can influence the velocity distribution of the entire population of the MW satellites~\citep{2016MNRAS.461.2212J}.

The current work is an extension of an analysis of the solar apex behaviour,
which is vital for studying the motion of galaxies outside the Local Group \citep{1996AJ....111..794K, 2008ApJ...676..184T, 2020AJ....159...67K}. 
% 1996AstL...22..455K, 1977ApJ...217..903Y, 1985ApJ...297...27D
In particular, by making an analysis of the solar apex variations as a function of the galactocentric radius, we discovered its peculiar behaviour for the inner population of the MW satellite galaxies, which, as we show below, can be explained by the infall of the LMC.

The paper is structured as follows.  
In Section~\ref{sec:Sample}, we present a sample of the MW satellites. 
The current knowledge of the solar apex with respect to the MW is given in Section~\ref{sec:SolarApex}.
Sections~\ref{sec:Dipole} and \ref{sec:Quadrupole} describe the solutions for the dipole and quadrupole terms of the observed velocity field, respectively.
In Section~\ref{sec:TwoPopulations}, we discuss the existence and properties of two kinematically different populations of the satellites. 
The model of velocity field perturbation by the motion of a massive satellite is described in Section~\ref{sec:LMCinfluence}.
We conclude in Section~\ref{sec:Summary}.

\section{Satellites of the Milky Way}
\label{sec:Sample}

\begin{figure}
\centering
\includegraphics[height=\columnwidth, angle=270, bb=15 95 357 680, clip]{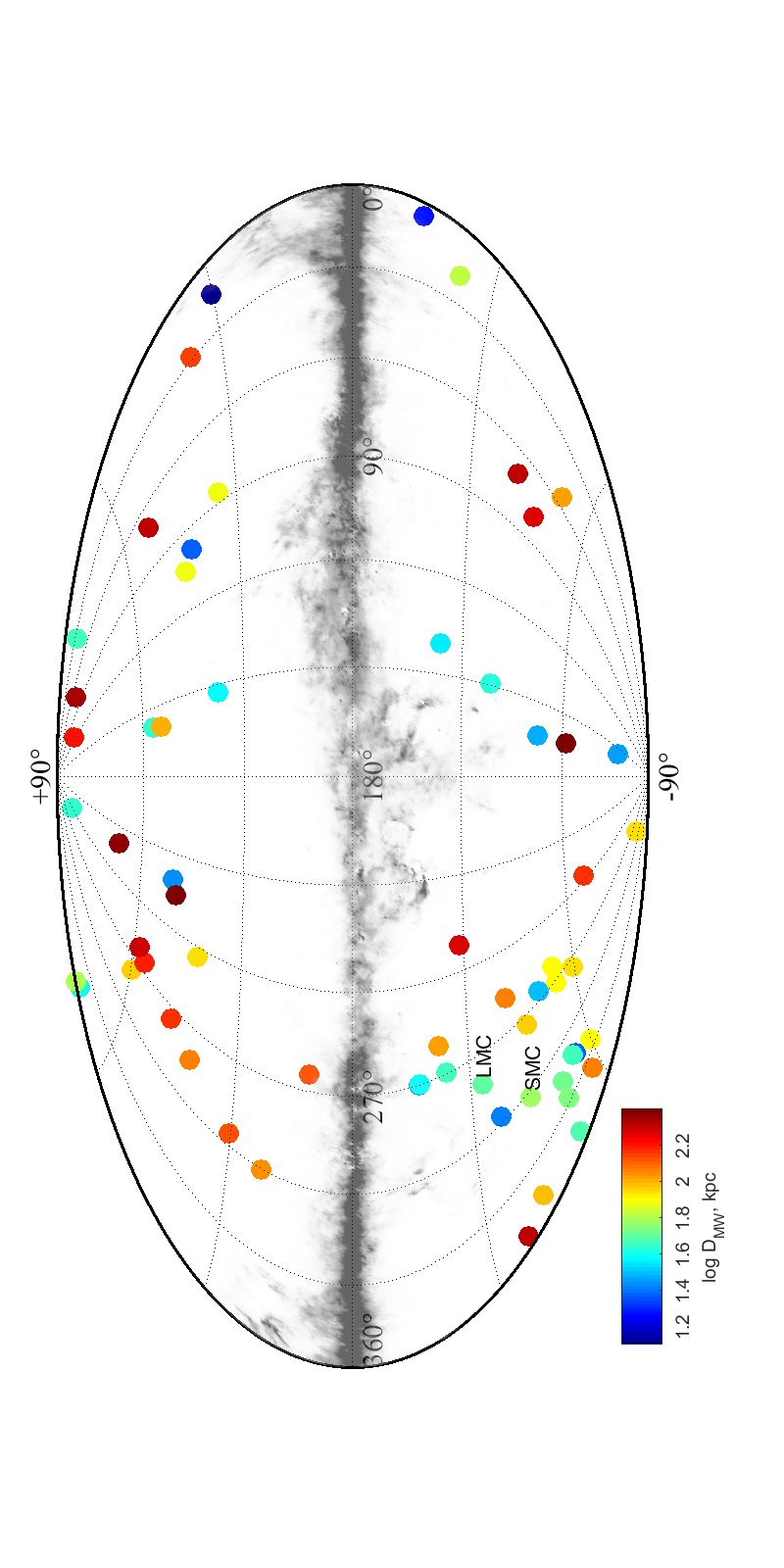}
\includegraphics[height=\columnwidth, angle=270, bb=15 95 357 680, clip]{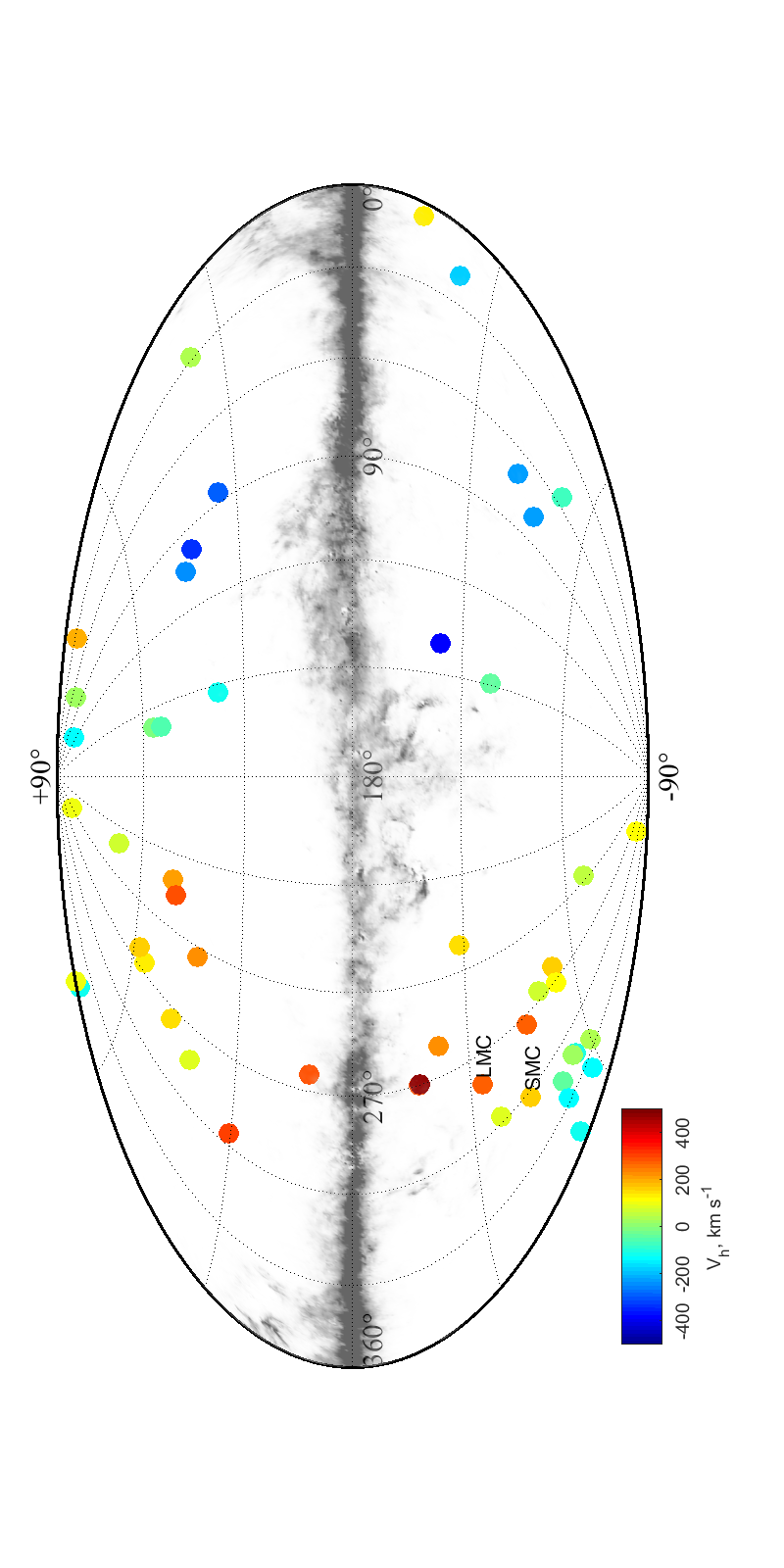}
\caption{
Distribution of the MW satellites over the sky in galactic coordinates. 
The Zone of avoidance in the MW is shown in gray-scale.
The direction towards the Galactic anticentre is in the centre of the map.
The upper panel illustrates the colour-coded distribution of the satellites by galactocentric distance, while the bottom panel shows their heliocentric radial velocities.
}
\label{fig:SkyMap}
\label{fig:VelocitySkyMap}
\end{figure}

In order to investigate the kinematics of the MW satellites, 
we selected 66 currently known dwarf galaxies within 450~kpc around our Galaxy. 
The data were collected from several recent studies of cosmic flows in the vicinity of massive galaxies \citep{2018A&A...609A..11K} 
including a systematic search for ultra-faint satellites in the Dark Energy Survey (DES) and Pan-STARRS1 (PS1) surveys~\citep{2020ApJ...893...47D}.
Our catalogue also includes the data from the works on proper motions of dwarf galaxies around the MW~\citep{2020AJ....160..124M,2022ApJ...940..136P}.
The final catalogue is presented in Table~\ref{tab:SatelliteList}, 
where we list the sky coordinates, distance and heliocentric line-of-sight velocity. 
The distribution of galaxies in the sky is shown in Fig.~\ref{fig:SkyMap}. 

\begin{figure}
\centering
\includegraphics[width=\columnwidth]{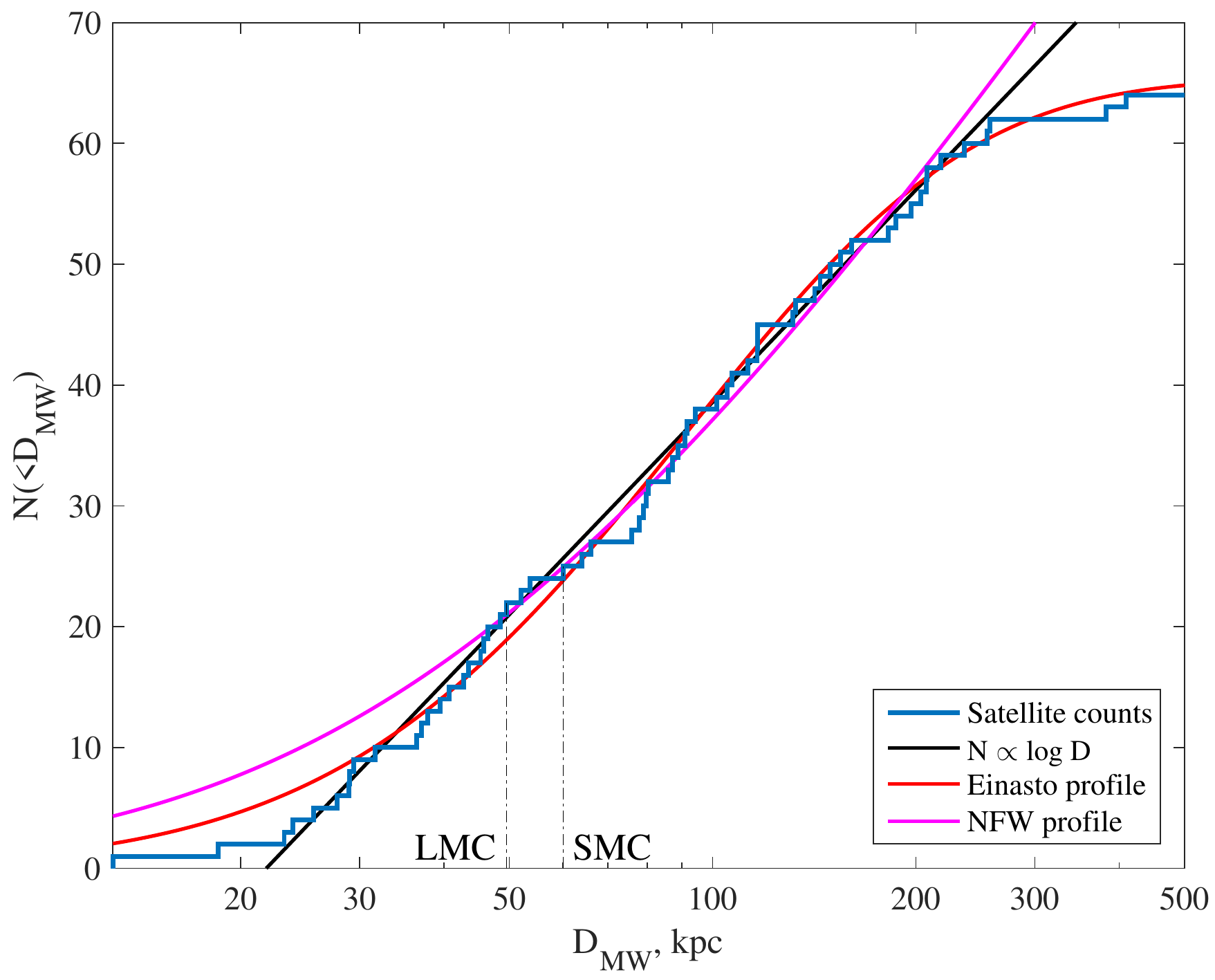}
\caption{
Cumulative number distribution of the satellites as a function of distance from the Galactic centre~(blue line). 
The linear fit of the distribution is shown by the black solid line,
while red and magenta lines correspond to Einasto and NFW profiles, respectively.
}
\label{fig:CumSum}
\end{figure}

\begin{figure}
\centering
\includegraphics[width=\columnwidth]{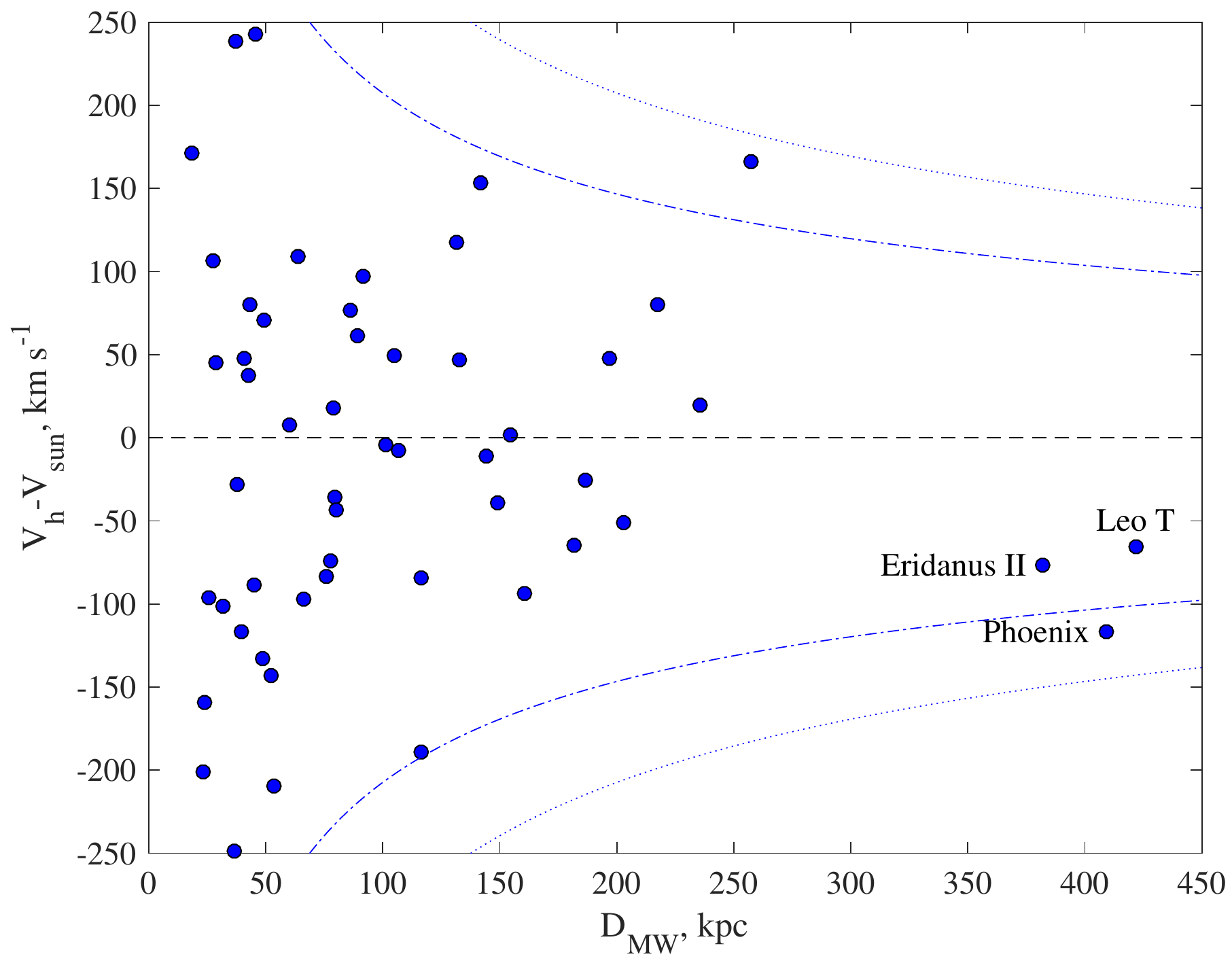}
\caption{
Velocity-galactocentric distance distribution of nearest galaxies.
The line-of-sight velocity is corrected for the motion of the Sun in the MW.
The dot-dashed and dotted lines correspond to circular and escape velocities for a point mass of $1\times10^{12}$~\Msun, correspondingly.
}
\label{fig:DistVel}
\end{figure}

We show the distance distribution of dwarf galaxies around the MW in Fig.~\ref{fig:CumSum}. 
Leaving out the completeness of the sample of satellites, which is outside the scope of this work, 
we find that the number of objects linearly increases as a function of $\log$-distance.
This relation corresponds to a decrease in the spatial number density of satellites as the cube of the distance, $n \propto r^{-3}$. 
The relation starts from Tucana~III, $\DMW\approx23$~kpc, 
which roughly corresponds to the boundary of the Galactic \textit{stellar} halo, $\approx25$~kpc~\citep{2016ARA&A..54..529B}, 
and continues out to Leo~I, $\DMW\approx260$~kpc, 
which roughly corresponds to the outer boundary of the MW \textit{dark matter} halo, $292\pm61$~kpc~\citep{2020MNRAS.496.3929D}. 
The number density distribution can be well described by the Einasto profile~\citep{1965TrAlm...5...87E}, 
$n \propto \exp(-0.82 r^{0.46})$, 
except for the innermost parts of the MW stellar halo, $\DMW\lesssim30$~kpc
(the fit is shown by the red line in Fig.~\ref{fig:CumSum}). 
The observed lack of dwarf galaxies in the proximity of the disc is likely caused by their disruption in the tidal field of the MW. 
The NFW profile~\citep{1996ApJ...462..563N} overpredicts the number of galaxies at very large distances from the MW, $\DMW\gtrsim260$~kpc,
however, it works well at smaller distances down to the Large Magellanic Cloud (LMC) at $\sim50$~kpc
(the fit is shown by the magenta line in Fig.~\ref{fig:CumSum}).

Beyond 260~kpc from the MW, we notice a gap in the distribution of satellites,
clearly visible in the velocity-distance diagram (Fig.~\ref{fig:DistVel}).
The next dwarf galaxies appear only at a distances of about $\DMW\approx400$~kpc.
They also differ in their kinematics.
Inside 260~kpc, satellites are very well randomised around MW in terms of velocities.
Whereas at distances of about 400~kpc, all galaxies have negative radial velocities.
This indicates that Eridanus-II, Phoenix and Leo~T have just begun their first fall into the halo of the MW.
From this we can conclude that the virial zone of our Galaxy extends at least up to 260, but no further than 380~kpc.
Therefore, for further analysis, we use only galaxies within 260~kpc of the MW.

Most of the radial velocities in our sample were adopted from \citet{2020AJ....160..124M} 
with the latest updates from \citet{2022ApJ...940..136P} and from the original articles.
Several missing galaxies, namely SMC, LMC, Crater, Bootes~III, Sagittarius~dSph, have been added from the database of the Local Volume galaxies\footnote{\url{https://www.sao.ru/lv/lvgdb/}}~\citep{2012AstBu..67..115K}.
In total 79 per cent of the satellites in the virial zone of the MW (50 out of 63) have precise radial velocity measurements with a typical uncertainty less than $2$~\kms{}.
The lower panel of Fig.~\ref{fig:VelocitySkyMap} presents the distribution of the radial velocities of the satellites over the sky in galactic coordinates.
One can see that the distribution clearly shows a large-scale dipole structure
with negative velocities in the direction of the solar motion around the Galactic centre, $(l, b)\approx(90^\circ, 0^\circ)$
and positive velocities in the opposite direction, $(l, b)\approx(270^\circ, 0^\circ)$.

Although the distribution of galaxies in Fig~\ref{fig:SkyMap} shows the above-mentioned plane of satellites 
and the apparent concentration of objects in the direction of LMC/SMC system, 
the celestial sphere is covered with satellites quite well.
This makes it possible to study full three-dimensional collective motion of a system of the dwarf galaxies around the MW 
and to estimate the solar apex relative to these galaxies.

\section{Running solar apex}
\label{sec:RunningSolarApex}

\subsection{Motion of the Sun in the Galaxy}
\label{sec:SolarApex}

In our analysis, we adopt the motion of the Sun in the MW from the review by \citet{2016ARA&A..54..529B},
where the authors collect and systematised information about the structure and properties of our Galaxy.
In particular, we assume that the Sun is located $8.2\pm0.1$~kpc away from the Galactic centre and $25\pm5$~pc above the plane of the Galaxy.
The solar motion relative to the Local Standard of Rest (LSR) is $(U_{\odot}, V_{\odot}, W_{\odot}) = (10.0\pm1, 11.0\pm2, 7.0\pm0.5)$~\kms,
where $U_{\odot}$ corresponds to the motion towards the Galactic centre,
$V_{\odot}$ is the velocity in the direction of rotation, 
and $W_{\odot}$ is the component perpendicular to the disc plane in the direction of the North Pole of the Galaxy.
The angular motion of the Sun in the Galactic plane relative to the the compact radio source Sgr~A* associated with a supermassive black hole at the centre of the MW is $\Omega_{\odot}=30.24\pm0.12$~\kmskpc{}~\citep{2004ApJ...616..872R,2008IAUS..248..141R}. 
This results in the tangential velocity of the Sun of $V_{g,\odot}=248\pm3$~\kms.
Combining the above together, the components of the solar motion relative to the Galactic centre are $(U, V, W) = (10.0\pm1, 248\pm3, 7.0\pm0.5)$~\kms{}.
Since the collective motion of the LSR relative to the Galaxy rotation remains uncertain, $V_{\rm LSR}=0\pm15$~\kms{}, 
\citet{2016ARA&A..54..529B} estimated the circular velocity of the Galaxy in the solar vicinity to be $238\pm15$~\kms{}.

\subsection{Measurement of the solar apex}
\label{sec:Dipole}

\begin{figure}
\centering
\includegraphics[width=\columnwidth]{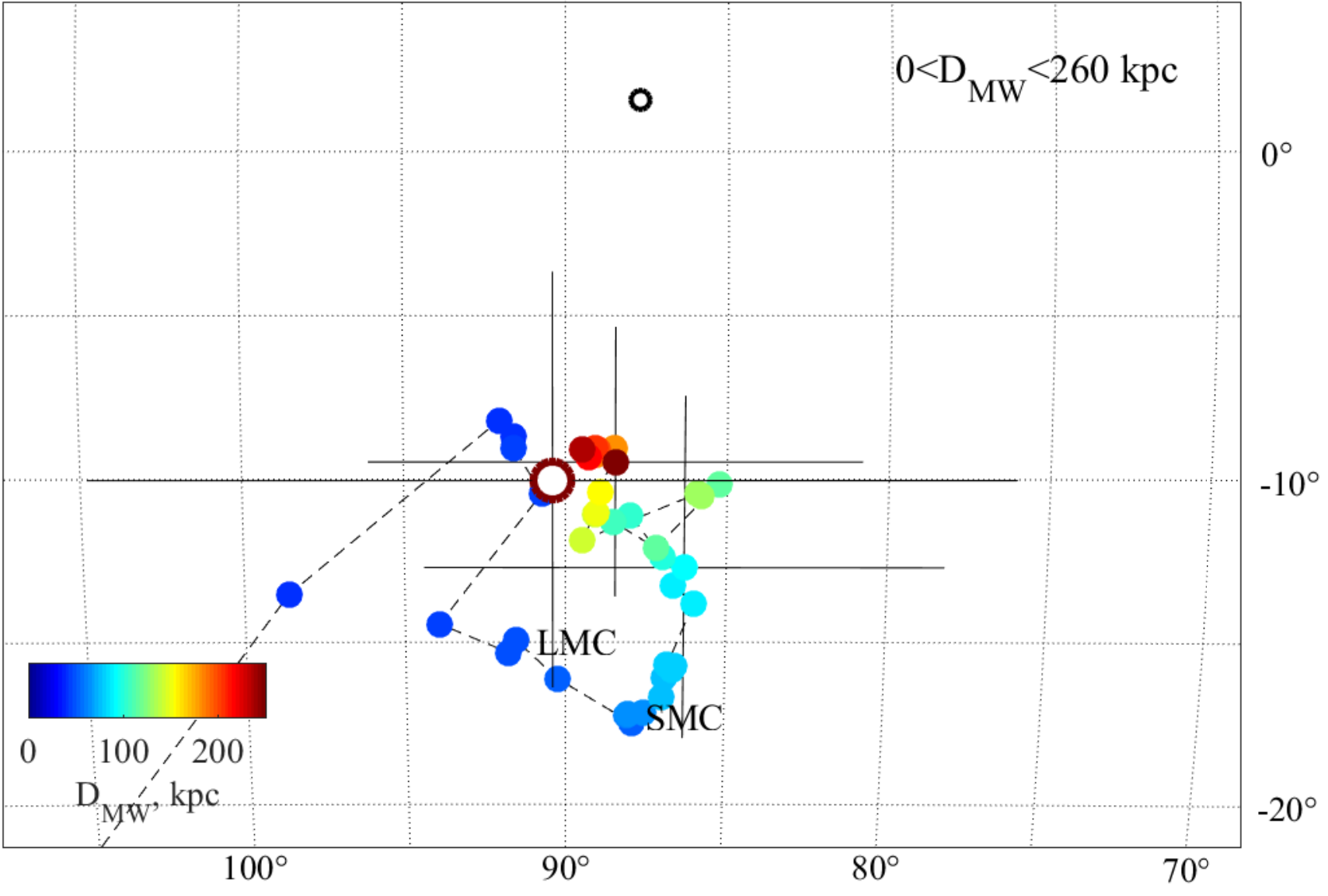}
\caption{
The `drift' of the running solar apex in the galactic coordinates.
The apex `trajectory' is presented by a dashed line.
The colour of the dots corresponds to the sample depth, namely the distance to the farthest galaxy in the sample.
The errors are shown only for two subsamples of 0--100 and 0--260~kpc.
The big brown open circle with errors corresponds to a subsample of galaxies between 100 and 260~kpc.
The small black open circle indicates the solar apex in our Galaxy.
Its size roughly corresponds to the measurement errors.
}
\label{fig:ApexMap:0-260}
\end{figure}

\begin{figure*}
\centering
 \includegraphics[width=\columnwidth]{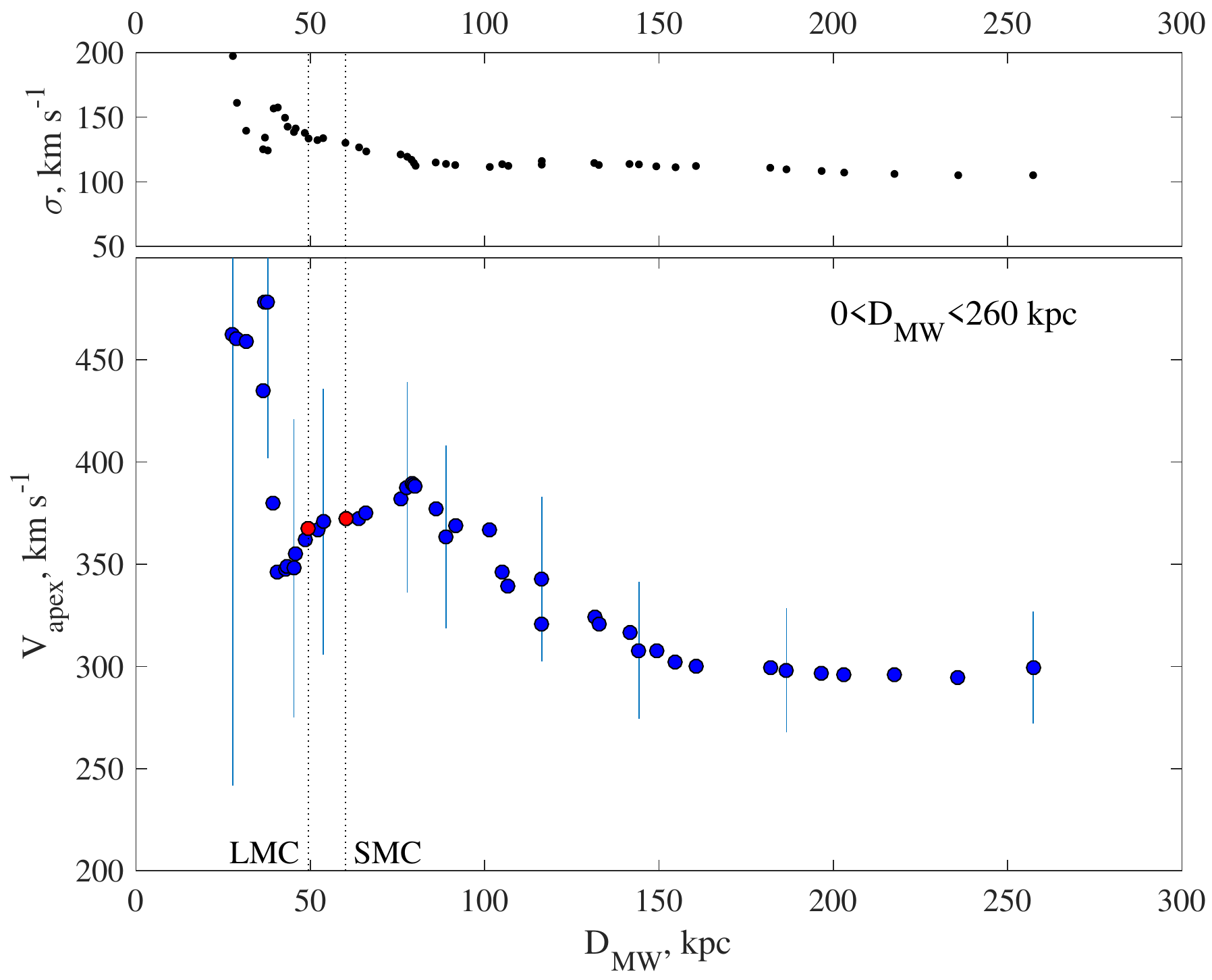}
\includegraphics[width=\columnwidth]{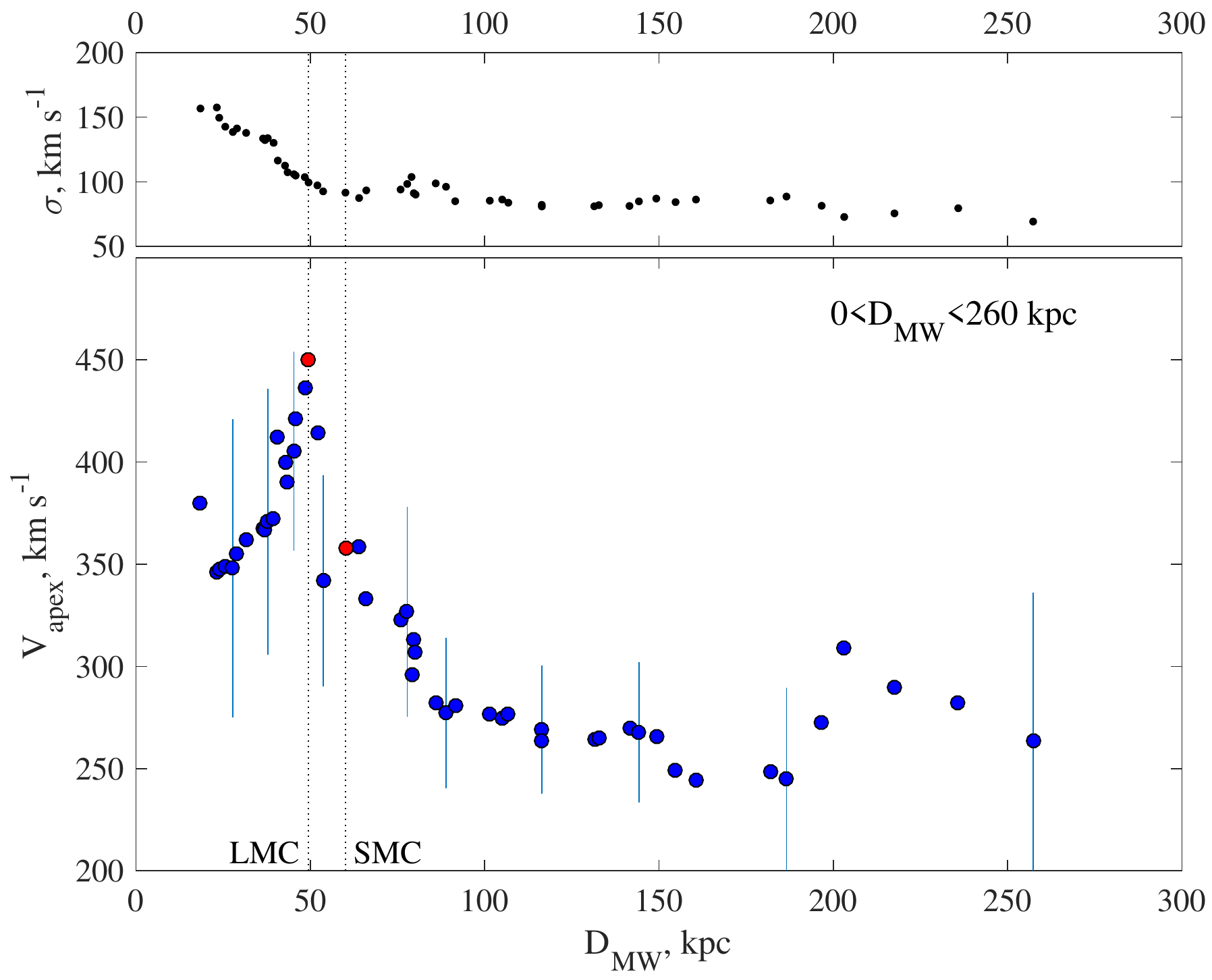}
\caption{
The `evolution' of the apex velocity with the galactocentric distance.
The standard deviation of the residual velocities are shown above.
Error bars are shown only for every fifth measurement, so as not to clutter the figure.
\textit{Left:} the running apex as a function of the satellite sample depth.
The samples containing the LMC and SMC for the first time are marked in red.
\textit{Right:} behaviour of the amplitude of the smoothed solar apex estimated from 21 satellites for each point.
The apexes centred on the LMC and SMC are also marked in red.
}
\label{fig:ApexVel}
\end{figure*}

The line-of-sight heliocentric velocity $\Vh$ of the MW satellites can be represented as:
\begin{equation}
\Vh = (\mathbfit{V}_{g} - \mathbfit{V}_{\odot})\cdot\mathbfit{n} + \varepsilon,
\end{equation}
where $\mathbfit{V}_{\odot}$ is the vector of the solar motion in the Galaxy,
$\mathbfit{V}_{g}$ is the velocity vector of a satellites relative to the Galaxy,
$\mathbfit{n}$ is the unit direction vector to a satellite,
and $\varepsilon$ is the random velocity measurement error.

Assuming that the motion of the satellites is uncorrelated,  we can consider the line-of-sight velocity of the galaxy, $\mathbfit{V}_{g}\cdot\mathbfit{n}$,  
as a random variable with a mean value equal to zero and a variance of the order of 100~\kms{}. In this case, the calculation of the solar apex is based on a linear regression problem:
\begin{equation}
\Vh =  -\mathbfit{V}_{\odot}\cdot\mathbfit{n} + \epsilon,
\label{eq:Dipole}
\end{equation}
where the model `error', $\epsilon$, is the sum of a random variable associated with the unknown direction of the galaxy motion and the velocity measurement error.
The one-sigma errors of the three components of the solar motion vector, $\mathbfit{V}_{\odot}$, are estimated in a standard way using covariance matrix and residual variance of the linear regression~(\ref{eq:Dipole}). 
For convenience, the vector of the Sun motion is converted into spherical galactic coordinates, and its errors are transformed in accordance with the propagation of uncertainties.
Note that the measurement errors of $\sim 1$--2~\kms{} are negligibly small 
compared to the velocity dispersion of satellites in the gravitational field of the MW.
At the same time, the solar apex, determined from the satellites, contains information not only about the motion of the Sun in the Galaxy, 
but also about the collective motions of the surrounding galaxies.
Since this model is purely focused on the analysis of velocity distribution, distance errors of the satellites do not affect the resulting solution.
Distance uncertainties can only slightly affect the composition of a subsample of galaxies due to variation of their distances.

Having sorted the satellites in order of increasing distance from the centre of the MW, 
we determine the solar apex depending on the current number of galaxies under consideration, $k$, 
and, in fact, on the distance from the centre of the Galaxy.
Despite the fact that each subsequent estimate is not statistically independent of the previous ones, 
this approach allows us to reveal trends in the behaviour of the solar apex.
In fact, this methodology repeats in the main details the analysis of the motion of our Galaxy relative to its neighbours in the Local Volume \citep{1996AJ....111..794K}.

The `drift' of the solar apex in the sky is shown in Fig.~\ref{fig:ApexMap:0-260}.
At small distances, the apex yaws across the sky and its amplitude jumps due to the small sample sizes.
Then points concentrate around the direction $(l,b)=(+86^\circ, -16^\circ)$ with the amplitude of $\sim390$~\kms{}.
After which there is a transition to a stationary value of $(l,b,V) = (+88^\circ, -9^\circ, 300~\kms{})$ for the entire sample of satellites within 260~kpc.
The apex drifts systematically (at the level of 2--2.5 sigma) below the expected direction of the motion of the Sun in the Galaxy.
The variation of the apex amplitude and the standard deviation of the residual velocities with sample depth is presented in the left panel of Fig.~\ref{fig:ApexVel}.
It is noteworthy that the apex velocity determined for nearby galaxies turns out to be systematically higher 
than for a more complete sample including distant satellites.
The transition between these two stationary points occurs at a distance of about 100~kpc, 
dividing the MW satellites into conditionally internal and external ones.

This feature becomes even more pronounced if we plot the running apex estimated for each satellite including 10 galaxies before and after it in the ordered list.
This approach is similar to data smoothing by the moving average filter.
As a result, each dot in the right panel of Fig.~\ref{fig:ApexVel} corresponds to the solar apex determined from the motion of 21 objects
(of course, we take into account only data that fall into the smoothing window and near the boundaries, the number of galaxies under consideration is less than 21, dropping down to 11 for the first and last galaxy in the sample).
The apex amplitude determined for internal galaxies below $\sim 80$~kpc is systematically higher than 300~\kms{}, reaching a maximum of 450~\kms{} for the LMC.
This value significantly exceeds the velocity of the Sun in the Galaxy of 248~\kms{} at the level of 4 sigma.
The smoothed apex above $\sim 80$~kpc sharply calms down and forms a plateau of $\sim270$~\kms{}.

A comparison of the two approaches shows that the internal galaxies at $\DMW<100$~kpc are responsible for both the systematic deviation of the apex for the full sample of satellites and their higher velocity dispersion.

\begin{table*}
\centering
\caption{The solar apex estimated from different samples of satellites of our Galaxy.}
\label{tab:Apex}
\begin{tabular}{rccrrrrrrrrrrrrr}
\hline\hline
\multicolumn{1}{c}{Sample} &
N &
$\sigma$ &
\multicolumn{1}{c}{$U$}	&
\multicolumn{1}{c}{$V$} &
\multicolumn{1}{c}{$W$} &
\multicolumn{1}{c}{$l$} & 
\multicolumn{1}{c}{$b$} & 
\multicolumn{1}{c}{$V$} &
\multicolumn{1}{c}{$t$} \\
&
&
\kms &
\multicolumn{1}{c}{\kms} &
\multicolumn{1}{c}{\kms} &
\multicolumn{1}{c}{\kms} &
\multicolumn{1}{c}{$^\circ$} & 
\multicolumn{1}{c}{$^\circ$} & 
\multicolumn{1}{c}{\kms} &
\\ 
\hline
\multicolumn{1}{c}{Sun}  &    &      & $+10 \pm1\phantom{0}$ & $+248 \pm3\phantom{0}$ & $  +7 \pm1\phantom{0}$ & $+87.7 \pm0.2$ & $ +1.6 \pm0.1$ & $248 \pm 3\phantom{0}$ &     \\
\hline

% ordinary LS
$  0<\DMW<100$           & 31 & 113 & $+23 \pm51$ & $+359 \pm44$ & $ -81 \pm33$ & $+86.3 \pm 8.1$ & $-12.7 \pm 5.3$ & $369 \pm44$ & 3.5 \\ 
$ 30<\DMW<85\phantom{0}$ & 22 & 101 & $+71 \pm54$ & $+414 \pm54$ & $-110 \pm35$ & $+80.2 \pm 7.4$ & $-14.7 \pm 4.8$ & $435 \pm53$ & 4.3 \\ 
$ 35<\DMW<81\phantom{0}$ & 21 & 100 & $+82 \pm54$ & $+430 \pm55$ & $-106 \pm34$ & $+79.3 \pm 7.0$ & $-13.6 \pm 4.6$ & $450 \pm54$ & 4.5 \\ 
$ 30<\DMW<100$           & 25 & 105 & $+62 \pm54$ & $+384 \pm48$ & $ -79 \pm32$ & $+80.8 \pm 7.9$ & $-11.5 \pm 4.8$ & $397 \pm48$ & 3.7 \\[5pt] 
$  0<\DMW<260$           & 50 & 105 & $ +8 \pm39$ & $+295 \pm28$ & $ -49 \pm21$ & $+88.5 \pm 7.6$ & $ -9.5 \pm 4.1$ & $300 \pm27$ & 3.0 \\ 
$ 30<\DMW<260$           & 44 & 101 & $+22 \pm41$ & $+295 \pm28$ & $ -45 \pm21$ & $+85.8 \pm 8.0$ & $ -8.6 \pm 4.0$ & $300 \pm28$ & 2.8 \\[5pt]
$100<\DMW<260$           & 19 & \phantom{0}84 & $ -2 \pm62$ & $+245 \pm37$ & $ -43 \pm27$ & $+90\pm14\phantom{.}$ & $-10.0 \pm 6.4$ & $249 \pm37$ & 1.7 \\ 

%\hline
%GC \phantom{$\DMW>10$} & 154 & 128 & $ +7 \pm13$ & $+149 \pm25$ & $ -11 \pm26$ & $+87.3 \pm 4.9$ & $ -4 \pm10\phantom{.}$ & $150 \pm25$ & 4.0 \\ 
%GC $\DMW>10$           &  48 & 134 & $ +6 \pm33$ & $+177 \pm37$ & $  +1 \pm32$ & $+88 \pm 11\phantom{.}$ & $ +0 \pm10\phantom{.}$ & $177 \pm37$ & 1.9 \\ 

\hline\hline
\end{tabular}
\end{table*}

The solar apex, estimated only from a sample of external satellites, $100<\DMW<260$~kpc, outside the `transition zone', 
is $(l, b, V) = (90\pm14^\circ, -10\pm6^\circ, 249\pm37~\kms{})$ with a standard deviation of the residual velocities of 84~\kms{}.
This value is in good agreement with the known motion of the Sun in the Galaxy.
The position of the solar apex for the external satellites is marked in Fig.~\ref{fig:ApexMap:0-260} by a big brown open dot with error bars.

The results are summarised in Table~\ref{tab:Apex}.
It contains information about
1) the selection criteria for satellites in the sample;
2) the number, N, of galaxies in the sample;
3) the standard deviation, $\sigma$, of the residual velocities of the model~(\ref{eq:Dipole});
4, 5, and 6) the spatial, $U$, $V$, $W$, components of the solar motion vector;
7, 8 and 9) the solar apex, $(l, b)$, in Galactic coordinates and its velocity, $V$;
and 10) $t$-test characterising the significance of the difference between the apex and the motion of the Sun in the Galaxy.
For convenience, the first line of the table shows the apex of the solar motion in the Galaxy from the review by \citet{2016ARA&A..54..529B}
described in the Section~\ref{sec:SolarApex}.

\subsection{Quadrupole term of the radial velocities}
\label{sec:Quadrupole}

The tidal influence of neighbouring galaxies should appear in the quadrupole term of the velocity distribution of the satellites.
The corresponding model can be written in the matrix form as:
\begin{equation}
\Vh = - \bm{\mathsf{n}}^{T} \bm{\mathsf{V}}_{\odot} + \bm{\mathsf{n}}^{T} \bm{\mathsf{H}} \bm{\mathsf{n}} + \epsilon,
\label{eq:Quadrupole}
\end{equation}
where $\Vh$ is an observed heliocentric velocity of the satellite,
$\bm{\mathsf{V}}_{\odot}$ is the solar motion vector,
$\bm{\mathsf{n}}$ is an unit vector in the direction to the satellite,
$\bm{\mathsf{H}}$ is a symmetric matrix, $(H_{ij}= H_{ji})$, with zero trace, $\sum H_{ii} = 0$, 
describing the quadrupole anisotropy of the velocity field,
and $\epsilon$ is a random component of the velocity.

We found the joint solutions for the solar apex and the quadrupole term of the velocity field 
for a complete sample of galaxies, $0<\DMW<260$, 
for a subsample of internal, $\DMW<100$, and external, $\DMW>100$, satellites, 
with and without taking into account the nearest galaxies, $\DMW<30$~kpc, 
that are most affected by the MW.
None of the considered samples has the significance of the quadrupole term exceeding the 3 sigma level.
The most significant solution, $\approx2.8$~sigma, is for the subsample of 31 inner satellites, $0<\DMW<100$~kpc.
However, the exclusion of the nearest galaxies, $\DMW<30$~kpc, from consideration completely destroys the quadrupole term 
and its difference from zero becomes miserable, 1.1 sigma.
Thus, we can conclude that the largest deviations from the dipole distribution are associated with the motion of the 6 nearest satellites.
The dynamics of these galaxies may be affected by the strong tidal influence of our Galaxy and 
by the dynamical friction in the halo of the MW.

\section{Two populations of satellites}
\label{sec:TwoPopulations}

\begin{table}
\centering
\caption{Additional dipole term with respect to the solar motion in the Galaxy.}
\label{tab:GalaxyApex}
\begin{tabular}{rccrrrr}
\hline\hline
\multicolumn{1}{c}{Sample} &
N &
\multicolumn{1}{c}{$l$} & 
\multicolumn{1}{c}{$b$} & 
\multicolumn{1}{c}{$V$} &
\multicolumn{1}{c}{$t$} \\
&
&
\multicolumn{1}{c}{$^\circ$} & 
\multicolumn{1}{c}{$^\circ$} & 
\multicolumn{1}{c}{\kms} &
\\ 
\hline

% ordinary LS 
$  0<\DMW<100$           & 31 & $ +83 \pm 26$ & $-38 \pm15$ & $142 \pm40$ & 3.5 \\ 
$ 30<\DMW<85\phantom{0}$ & 22 & $ +70 \pm 17$ & $-34 \pm11$ & $212 \pm49$ & 4.3 \\ 
$ 35<\DMW<81\phantom{0}$ & 21 & $ +69 \pm 16$ & $-30 \pm10$ & $226 \pm50$ & 4.5 \\ 
$ 30<\DMW<100$           & 25 & $ +69 \pm 21$ & $-31 \pm13$ & $169 \pm45$ & 3.7 \\ 

\hline\hline
\end{tabular}
\end{table}

The analysis of the radial velocities of the MW satellites suggests the existence of two kinematically different populations.

The external satellites (19 objects) located further than 100~kpc show a random distribution of radial velocities 
with respect to our Galaxy with a dispersion of 84~\kms{}.
The dipole term of the velocity distribution is in reasonably good agreement with the motion of the Sun in the MW.
This is consistent with the standard concepts of galaxy group kinematics and modern cosmological modeling.

On the other hand, the apex determined from the nearby satellites within 100~kpc demonstrates 
a significant deviation from the known solar motion in the Galaxy.
Table~\ref{tab:GalaxyApex} summarises this difference for four samples of internal satellites.
The additional dipole is most pronounced at 4.5 sigma level in the kinematics of 21 galaxies located between 35 and 81~kpc from the Galaxy.
It seems that the MW flies through a swarm of internal satellites 
at a speed of $226 \pm50$~\kms{} in the direction of $(l, b) = (+69\pm16^\circ, -30\pm10^\circ)$.
This result seems surprising, 
because the satellites within 100~kpc should be virialized and 
the probability of preservation of the coherent motions is small.

\subsection{First passage}
\label{sec:FirstPassage}

In fact, we expect the correlated motion to persist only on the first passage of galaxies through the halo of the MW.
This scenario is supported by a recent work in which \citet{2021ApJ...922...93H} analysed 3D velocities, angular momentum, and total energy of 40 MW satellites using the \textit{Gaia} data.
They argue that nearby dwarfs, $<60$~kpc, are not long-lived satellites, 
but entered the MW group $\le2$ Gyr ago and are on their first flyby.
This is also consistent with the stellar population of these dwarf galaxies showing a small fraction of young stars.
The LMC is the most prominent galaxy passing the perigalacticon for the first time \citep{2007ApJ...668..949B} and entered within the virial zone about 1--3~Gyr ago.
Together with its escort \citep{2016MNRAS.461.2212J}, LMC may create a significant effect of the bulk motion with an amplitude of about 200~\kms.
It is worth to note that, using the proper motions of 52 dwarf spheroidal galaxies,
\citet{2022ApJ...940..136P} found no evidence that most satellites lie near their pericentre.

Figure~\ref{fig:ApexVel} suggests that the features of the kinematics of the internal satellites are associated with the motion of the LMC.
Indeed, the maximal bulk motion of the satellites is revealed when considering a layer of 21 galaxies within $35 \lesssim \DMW \lesssim 80$~kpc centred on the LMC.

\begin{figure}
\centering
\includegraphics[width=\columnwidth]{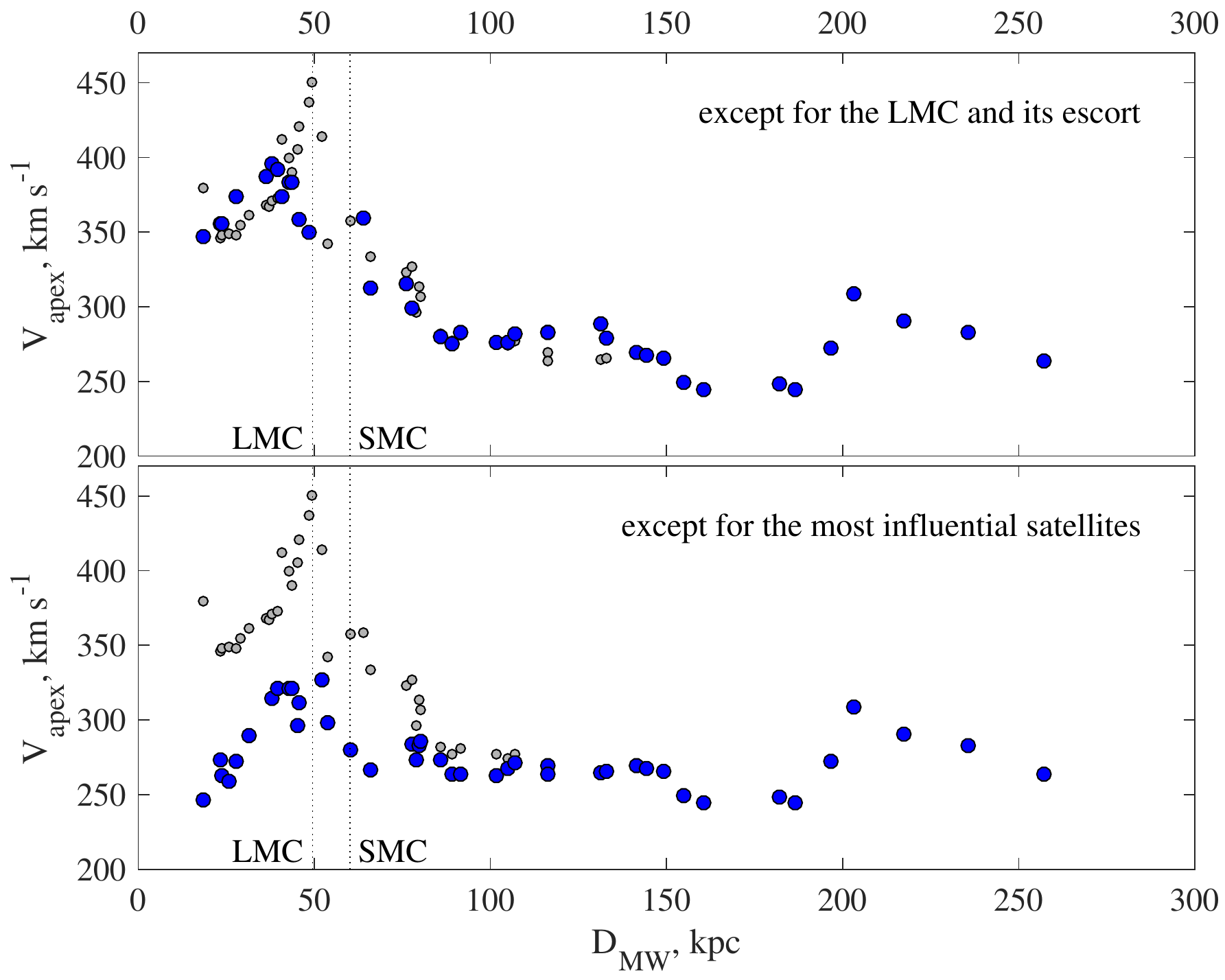}
\caption{
Behaviour of the amplitude of the smoothed apex after the elimination of some satellites.
Small light gray points correspond to the full sample, as shown in the right panel of Fig.~\ref{fig:ApexVel}.
\textit{Top}: the LMC and its most probable 12 members are excluded from consideration.
\textit{Bottom}: the 8 most influential galaxies are excluded. The top panel shows that the effect of the infall of the LMC-connected group of satellites can not explain the observed running solar apex variations.
}
\label{fig:SmoothApexNoLMC}
\end{figure}

However, a simple test for excluding from consideration the LMC and all escort dwarfs with close velocities, distances and positions in the sky, namely Tucana~IV, Pictor~II, Tucana~V, Tucana~II and SMC, reduces the amplitude of the peculiar motion, but does not change the picture qualitatively.
The additional bulk flow of the remaining 26 satellites within $\DMW=100$~kpc is $136\pm44$~\kms.
With the exclusion of Hydrus~I, Carina~III, Reticulum~II, Carina~II, Horologium~II, Horologium~I and Phoenix~II as the most probable members of the LMC group, according to \citet{2016MNRAS.461.2212J} and \citet{2018ApJ...867...19K}, the significance of the bulk motion inside $\DMW<100$~kpc drops to the level of 2.4 sigma ($V=146\pm62$~\kms{} from 19 galaxies).
This case is shown in the top panel of Fig.~\ref{fig:SmoothApexNoLMC}.
If we also exclude the potential LMC members Tucana~III, Draco~II, Grus~II, Reticulum~III, Grus~I, Pictor~I and Hydra~II, then the significance of the collective motion decreases even more, but only 15 galaxies remain in the sample of the internal satellites.

\begin{figure*}
\centering
\includegraphics[width=1\hsize]{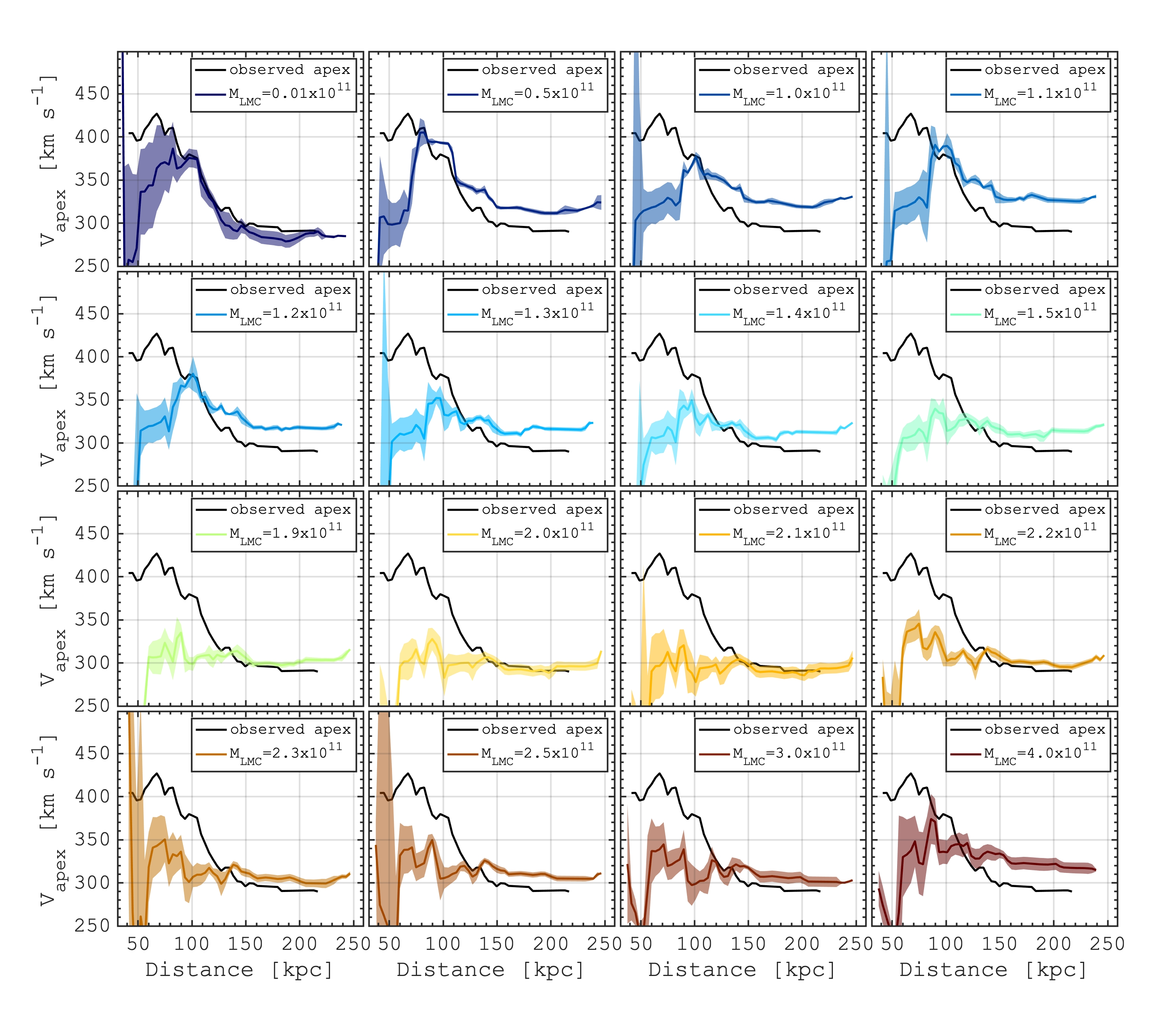}
\caption{
Solar apex based on the MW satellites orbits integration. Black curves are the same across the panels and show the present-day Solar apex. Coloured lines in different panels show the apex variation calculated after the integration of the satellites orbits, $5$~Gyr backwards and then forward to the present day. In the backward integration, we assume a given mass of the LMC; after that, we remove the LMC mass from the system and integrate all the satellites orbits forward. This approach allows testing the contribution of the massive LMC to observed kinematics of the MW satellites. One can see that in case of massive LMC, $\approx 1.4$--$2.2\times 10^{11}$~\Msun{}, the inner apex feature~($<100$--120~kpc) is diminished suggesting its causation by the LMC.
}
\label{fig:apex_back_forw}
\end{figure*}

\begin{figure*}
\centering
\includegraphics[width=1\hsize]{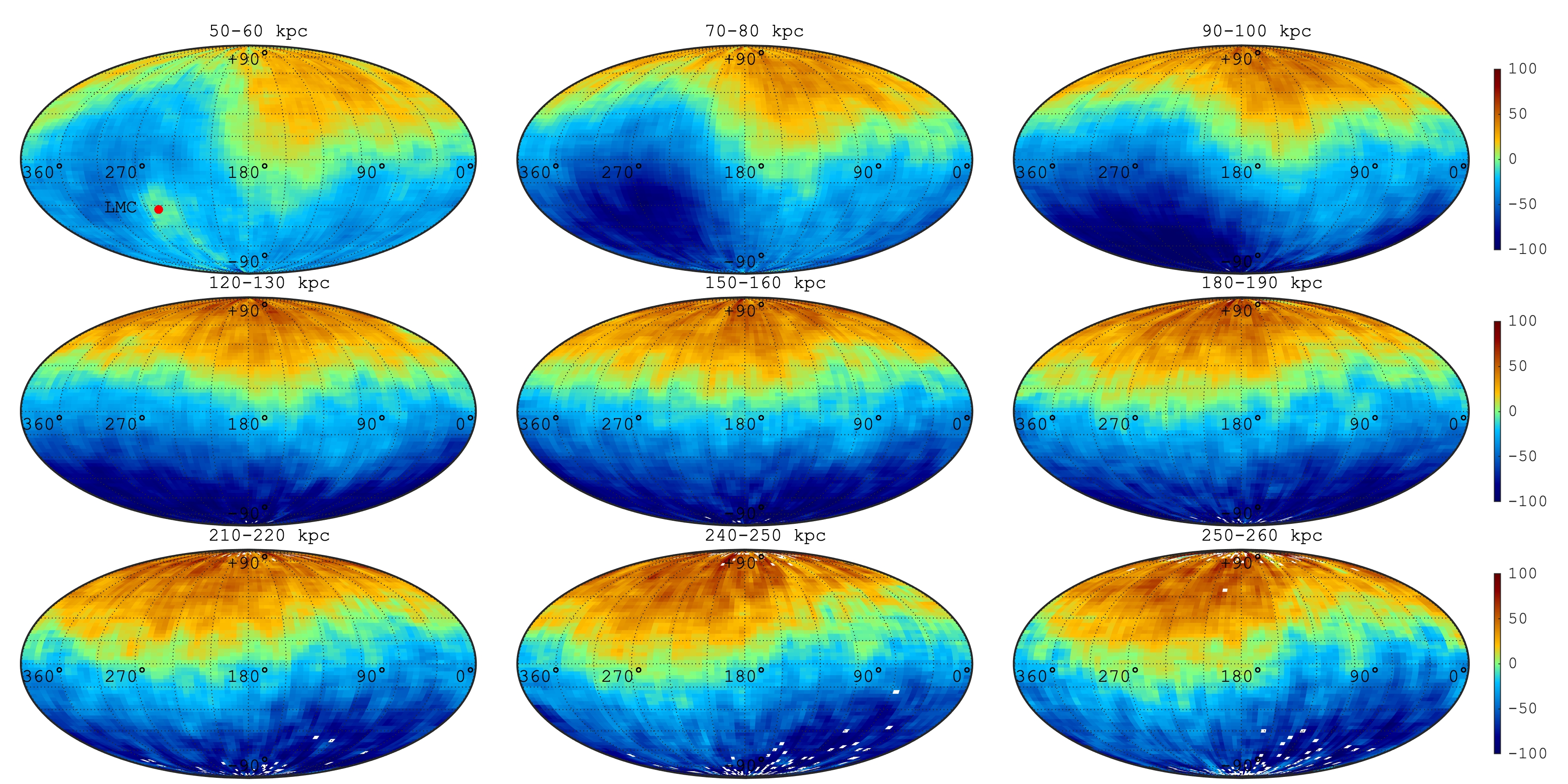}
\caption{
Sky-projections~($l,b$) of the mean line-of-sight radial velocity of the MW halo particles in the $N$-body simulations of the MW-LMC interaction. Each panel corresponds to the velocity distribution averaged for particles in a given galactocentric range, as stated in each panel. In this simulation, the total mass of the LMC is $2\times 10^{11}$~\Msun and its location is shown in the left top panel by the red circle.
}
\label{fig:model_velocity_maps}
\end{figure*}

Another way is to find the satellites most responsible for the deviation from the solar apex.
For this purpose, we search for a galaxy whose exclusion from consideration provided the greatest suppression of the dipole additional to the solar one.
Having eliminated this galaxy, we iteratively repeat the previous step until the maximum significance of the deviation falls below $t=2$.
The resulting behaviour of the smoothed apex velocity is shown in the bottom panel of Fig.~\ref{fig:SmoothApexNoLMC}.
The biggest contribution to the deviation of the apex from the solar one is made by Carina~II. 
Its exclusion reduces the significance of the additional dipole from $t=4.5$ to 3.6.
The galaxy with the highest negative velocity, Triangulum~II, is the second most influential satellite.
Together with Bootes~I they provide the significance of the deviation from the solar apex above 3 sigma.
Further exclusion of Draco, LMC, Segue~2, Carina~III and Grus~II makes the significance of the deviation less $t=2$.
It is important to emphasise that 4 of these 8 galaxies are members of the LMC group according to \citet{2016MNRAS.461.2212J} and \citet{2018ApJ...867...19K}.
This supports the hypothesis that the behaviour of the apex is associated with the first flyby of a group of galaxies around the MW.
On the other hand, the half of these most influential galaxies are not directly associated with the LMC, 
and this may indicate that the LMC perturbs their motion.

\subsection{Possible rotation of the inner satellites of the MW}
\label{sec:PlaneRotation}

A possible explanation for the observed collective motion of the internal satellites is the global rotation of this subsystem around the Galaxy.
This will create a pattern in the velocity field that mimics a dipole distribution, 
but it must depends on the distance to the objects.
As a consequence, the dipole solution will differ significantly from the apex associated with the solar motion in the MW.
Note that, due to projection effects, the component of the specific angular momentum pseudovector is directed to the centre of rotation 
does not make any contribution to the radial velocity field.
Therefore, the maximum and minimum of rotation will always be perpendicular to the direction to the centre of the Galaxy, 
in other words, they will have a galactic longitude of 90 or 270 degrees.
The observed orientation of the dipole additional to the solar apex (see Table~\ref{tab:GalaxyApex})
has a longitude of 70--80, which is consistent with the assumption about the rotation of the subsystem of satellites.

This hypothesis finds confirmation in the study of satellite planes.
Using the proper motions of 11 classical satellites of the MW, 
\citet{2013MNRAS.435.2116P} found a coherent orbital alignment for most (from 7 to 9) of the considered satellites.
This indicates that a vast polar structure of satellites forms a rotationally stabilised plane around the Galaxy.
In addition, \citet{2013Natur.493...62I} showed that the planar structure of satellites around the Andromeda galaxy rotates about the host galaxy.
These discoveries convincingly show that a significant number of satellites have the same dynamical orbital properties and direction of angular momentum.
Thus, the coherent rotation of satellite subsystems can be a widespread phenomenon.

A simple model of the radial velocity field, which takes into account the rotation of the system with a constant circular velocity, 
can be described by the equation:
\begin{equation}
V = \mathbfit{S}\times\mathbfit{n}_{\rm MW}\cdot\mathbfit{n}_{\odot} + \epsilon,
\end{equation}
where $V$ is the radial velocity of the galaxy corrected for the solar apex in the Galaxy;
for the spin pseudovector, $\mathbfit{S}$, only two tangential components can be determined;
$\mathbfit{n}_{\rm MW}$ and $\mathbfit{n}_{\odot}$ are unit vectors directed towards the galaxy from the MW centre and from the Sun, respectively.
Taking into account properties of the scalar triple product, 
it is easy to show that the observed peculiar component of the velocity field associated with the rotation of the system 
decreases inversely with the distance to the galaxy, $V\propto S/\DMW$.
As an result, even small peculiar velocity for the distant galaxies could require an high rotation speed.
This model can describe the velocity field quite well.
The standard deviation of the residuals of 105 for the $30<\DMW<85$~kpc sample is only slightly higher than 101~\kms{} for the dipole solution.
However, the model requires an incredibly high circular velocity of $970\pm260$~\kms{}, which cannot be physically justified.

Nevertheless, it is worth to note that tangential velocities, not included in our analysis, provide key information to distinguish rotation from dipole motion of the satellites, as stressed by \citet{2021NatAs...5..251P}.

\section{Impact of the LMC on the solar apex}
\label{sec:LMCinfluence}

One possible explanation for the observed bulk motion of the internal satellites may be 
the influence of the LMC on the position of the Galaxy within its halo. Recent studies show that the total mass of the LMC should be on the order of 1--$2\times10^{11}$~\Msun, which is a significant fraction of the mass of the MW halo. The presence of such a close and massive neighbour will inevitably affect the dynamics of the entire system \citep{2015ApJ...802..128G,2021ApJ...923..140G}.
In particular, it is expected that the MW will be accelerated relative to its outskirts beyond 30~kpc.
Using a sample of about 500 stars in the distant halo of the Galaxy, 
\citet{2021MNRAS.506.2677E} found an asymmetry in the distribution of radial velocities,
which is consistent with the acceleration of the inner halo caused by the LMC.

In their studies, \citet{2020MNRAS.494L..11P,2021NatAs...5..251P} say that the LMC can affect our Galaxy in different ways. In particular, there is a shred of evidence that the MW disc is moving with respect to stellar tracers in the outer halo with the velocity of 32$\pm$4~\kms{}, in the direction, which points at an earlier location on the LMC trajectory. The resulting reflex motion is detected in the kinematics of outer halo stars and MW satellite galaxies.

We test a hypothesis that a massive galaxy moving through the MW halo can cause a prominent dipole-like velocity pattern of the satellites, similar to the one observed in the lower panel of Fig.~\ref{fig:VelocitySkyMap}. In particular, as it is stated above, we study the impact of the LMC on the line-of-sight velocity distribution of objects in the halo of the MW.

\subsection{Solar apex without the LMC-induced perturbation}

We start our exploration of the LMC impact on the observed variations of the solar apex by studying the orbits of the MW satellites. We use a technique similar to the one described in \citet{2022MNRAS.511.2610C}, where first we integrate the orbits of the MW satellites \citep[with full 6D phase-space information from][]{2022A&A...657A..54B} backwards in time taking into account dynamical friction of the MW affecting the LMC orbit and a reflex motion of the MW caused by the interaction with the LMC. Next, we use the positions and velocities of the satellites (and the MW) and integrate the entire system forwards till the recent time but removing the mass of the LMC. Therefore, after this backwards-forward integration, we are able to find the phase-space configuration of the MW satellites in the absence of the LMC, which induces a significant kinematic asymmetry in the satellite MW distribution \citep[see Fig.~7 in][]{2022MNRAS.511.2610C}.

In our models, the LMC is a single spherical NFW halo with the scalelength of $\rm 8.5\ kpc \times (\MLMC/10^{11} \Msun)^{0.6}$ truncated~at 10 scalelengths reproducing the observed rotation curve if the LMC mass is about $(1$--$3) \times 10^{11}$~\Msun~\citep{2021MNRAS.501.2279V}. We explore the LMC mass range of $(0.01$--$4) \times 10^{11}$~$\Msun$, where the lowest limit LMC mass aimed to test a lack of the LMC-induced perturbation on the line-of-sight kinematics of the MW satellites.

In the orbit integration, the MW galaxy is represented by a multi-component composite potential model, including a spherical DM halo, stellar bulge, thin and thick stellar discs with parameters adopted from \citet{2017MNRAS.465...76M}. Therefore, in our calculations, both the LMC and the MW galaxies are represented by the potentials, while the satellite galaxies are considered as massless test particles. The backward integration time is taken to be $5$~Gyr which places a massive LMC well beyond the virial radius of the MW. Finally, after subtraction of the MW reflex motion at $t=0$ the phase-space data about the MW satellites after the forward integration are used to calculate the solar apex.

In Fig.~\ref{fig:apex_back_forw}, we show the running solar apex calculated after the backwards-forward integration of the MW satellites orbits assuming different masses of the LMC, as indicated in each panel. For reference, in each panel, we show the observed Solar apex as a solid black line. The shaded area around the lines shows a 1-sigma scatter of the solution based on the kinematics of the MW satellites over the last 0.5~Gyr in the forward integration. One can see that once we remove the impact of the LMC, the behaviour of the solar apex changes significantly. In particular, in the models with $\MLMC>1.3\times 10^{11}$~\Msun, the inner satellites do not longer show the observed peculiarity. The Solar apex distribution is nearly flat in the case of $\MLMC\approx 2\times 10^{11}$~\Msun, which is close to some estimates of the LMC mass~\citep{2007ApJ...668..949B, 2006ApJ...652.1213K, 2013ApJ...764..161K, 2016MNRAS.456L..54P, 2021MNRAS.501.2279V, 2020MNRAS.495.2554E, 2019MNRAS.487.2685E} implying the first-infall scenario for the LMC. Note also that a larger mass of the LMC~($>2.2 \times 10^{11}$~\Msun) results in a more substantial perturbation of the MW satellites, which results in the appearance of peculiar apex distribution, which, thus, provides a good constrain on the LMC mass.

\subsection{Tracing the impact of the LMC on the apex in simulations}

\begin{figure*}
\includegraphics[width=1\hsize]{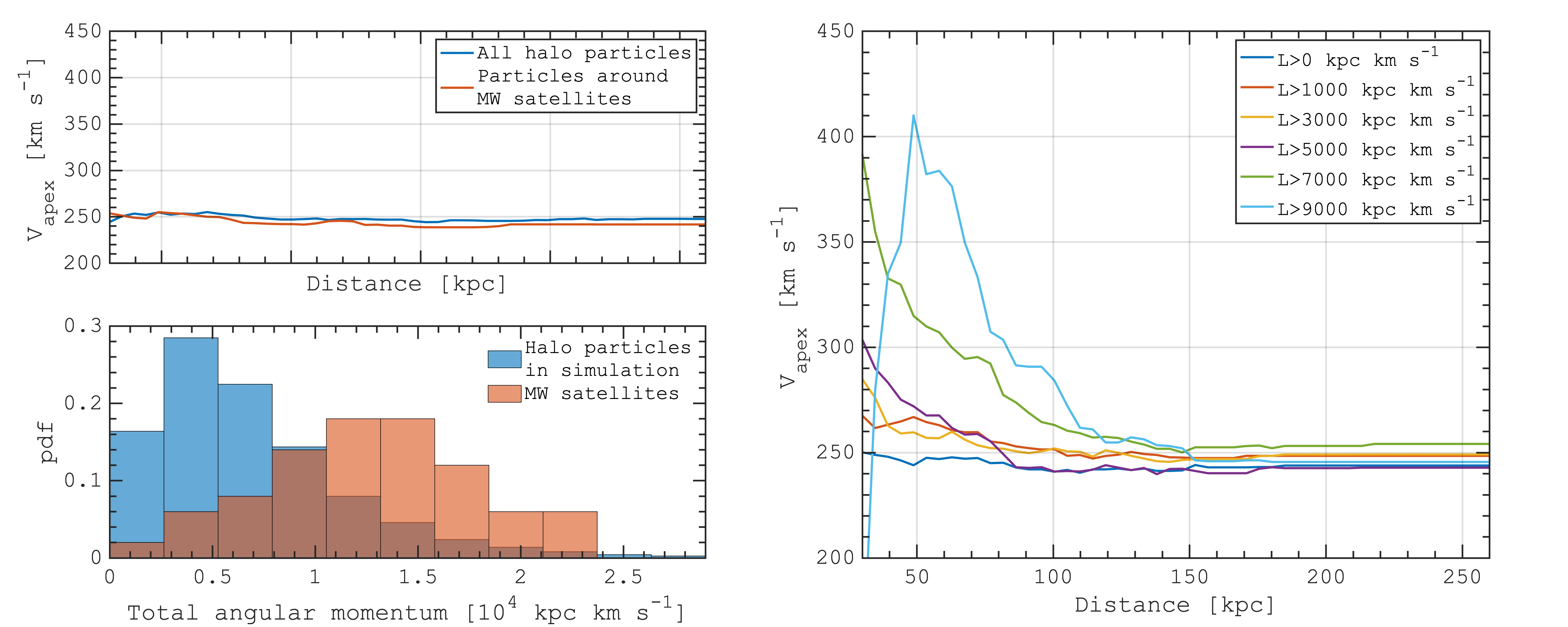}
\caption{
Solar apex calculation based on the halo particles in the $N$-body simulation of the MW-LMC interaction. {\it Left top:} solar apex calculated using all the DM particles~(blue) and particles which are located around the MW satellites~(red). The panel shows no manifestation of the LMC passage in the Solar apex despite a substantial kinematic response of the halo~(see Fig.~\ref{fig:model_velocity_maps}). {\it Left bottom:} comparison of the total angular momentum distribution of the halo particles in our $N$-body simulation~(blue) and the MW satellites~(red). The distributions suggest that the MW satellites show significant non-radial motions or their orbits are not radially hot, especially in comparison to the DM particles $N$-body simulations. {\it Right:} solar apex calculated using particles in the vicinity of the MW satellites also selecting particles with a total angular momentum larger than a given value. The figure shows that the observed Solar apex variations as a function of distance is seen among the particles with substantial angular momentum, suggesting a kinematic differentiation of the LMC impact in the MW halo populations.
}
\label{fig:Apex_AMdistr}
\end{figure*}

In order to understand better the dynamic response of the halo objects on the passage of the LMC galaxy, we simulate a dynamic evolution of the MW-type disc galaxy embedded into a massive DM halo with taking into account a massive object. In particular, we performed a set of high-resolution $N$-body simulations of the MW-LMC-like interactions. The initial conditions for these models are taken from our previous analysis. We use the positions of the MW and the LMC galaxy from the backward integration of the orbits 5~Gyr ago. Since both galaxies are represented by the analytic potential, we used them to generate the particles distribution using the AGAMA software~\citep{2019MNRAS.482.1525V}. In this case, our simulations take into account the direct effect of the LMC on the halo, as well as the reflex motion of the MW in response to the LMC.

In the simulations, the LMC is represented by $10^6$ particles; while the MW bulge, stellar thin and thick discs consist in $5\times 10^5$ each, and the DM halo of the MW is populated by $2\times 10^7$ particles. The latest allows us to trace the details of the kinematic response of the halo populations on the first infall of the LMC. The simulations were run using our parallel version of the \textsc{tree-grape} code~\citep{2005PASJ...57.1009F,2014JPhCS.510a2011K} with multithread usage under the SSE and AVX instructions providing us with a good performance and high-precision solutions of similar problems~\citep{2018MNRAS.481.3534S,2020A&A...638A.144K,2022A&A...663A..38K}. For the time integration, we used a leapfrog scheme with a fixed step size of 0.1~Myr. In the simulation, we adopted the standard opening angle $\Theta=0.5$. In total, we run eight simulations with a different mass of the LMC; however, here, we present a single model with $\MLMC=2\times 10^{11}\Msun$ since it provides the best explanation of the observed Solar apex behaviour and also in agreement with a number of recent works.  

In our model, we do not take into account the population of the satellite galaxies; however, the presence of the extended live DM halo allows us to use its particles to probe the kinematic response on the passage of the LMC-like massive satellite. Therefore, to compare the impact of the LMC passage with the velocity distribution of the satellite galaxies in the MW, we use the line-of-sight velocities of the DM particles. 

First, we take a look on the full-sky perturbation of the MW halo by the LMC. In Fig.~\ref{fig:model_velocity_maps} we show the mean line-of-sight velocity maps of the halo particles in our simulation at the time when the location of the LMC-like object corresponds to the present-day parameters. We map the mean radial velocity component in several bins of the galactocentric distance, as it is stated in each panel. In the figure, we observe a pattern previously reported in several recent works \citep{2019ApJ...884...51G,2020ApJ...898....4C,2020MNRAS.498.5574E}, which studied the motions of both the LMC and MW with different $N$-body simulations. In our simulation, the outer halo shows largely inward/outward motions relative to the MW mid-plane caused by the outer halo wake behind the LMC orbit~\citep{2019MNRAS.488L..47B, 2021ApJ...916...55T, 2021Natur.592..534C}. However, the velocity perturbation of the inner~(<100-120~kpc) galaxy is very complex and varies strongly with the position in the sky.

Since we have all the positions and radial velocities of the simulated MW halo particles, we calculate the solar apex distribution using all the particles in the halo and, since the radial velocity pattern is a function of the coordinates~$(l,b)$, (see Fig.~\ref{fig:model_velocity_maps}), we compare it to the signal based on the particles located in the close proximity of the MW satellites. In the left panel of Fig.~\ref{fig:Apex_AMdistr}, we show the solar apex distribution based on the entire halo~(blue) and particles selected in a $5^\circ$ circle around the MW satellites in the sky and located no further than 10~kpc~(red). Therefore, for the red curve, we expect to find behaviour similar to the one we observe in the MW. However, both curves in the left panel show no signal, suggesting an isotropic averaged motion of the halo particles in our model, even in the presence of the massive LMC. 

It is worth noticing that once we average the kinematic response, our results depend on the intrinsic properties of the tracer and its ability to `feel' the perturbation of the underlying gravitational potential~\citep{2020MNRAS.498.5574E, 2021ApJ...923..140G, 2022MNRAS.511.2610C, 2022ApJ...932...70P}. Therefore, the DM particles in our simulations may not have the same kinematics properties distribution as we see for the MW satellites. For instance, in the central panel of Fig.~\ref{fig:Apex_AMdistr}, we show the total angular momentum distributions~($L$) for the DM particles~(blue) and the MW satellites~(red). In our case, the DM is dominated by the radially-hot orbits with small angular momentum, while the MW satellites have less radial orbits, higher tangential velocities and, thus, higher angular momenta. 

In order to test whether the observed solar apex distribution depends on the kinematic properties of the halo tracers, we restrict our selection of the DM particles around the MW satellites to the particles which have a total angular momentum larger than a certain value. In the right panel of Fig.~\ref{fig:Apex_AMdistr} we show the solar apex calculated for different minimum values of $L = 1000$, 3000, 5000 and 9000~kpc\,\kms{} of the DM particles. 
The selection of tracer particles with substantial angular momentum results in high velocities of the apex at small galactocentric distances. Specifically, by choosing DM particles with $L>9000$~kpc\,\kms, the solar apex radial variations match the observed data.
This value of the angular momentum cut allows selecting particles which show the rotation more consistent with the MW satellites~(see middle panels of Fig.~\ref{fig:Apex_AMdistr}). The inner particles~($<100$--120~kpc) show a bulk motion relative to the Sun of about 100--150~\kms{} while the outermost distribution remains flat, which reproduces well our Fig.~\ref{fig:ApexVel}. Therefore, we suggest that since the MW satellites have substantial angular momentum or they are not on hot radial orbits, they are more affected by the LMC-induced perturbation. 

The effect of the differential response of populations with different kinematics on the gravitational potential perturbation, which we observe in the halo of the simulated MW, is known in the context of the stellar dynamics of galactic discs. In particular, \citet{2017MNRAS.469.1587D} showed that kinematically hot MW disc populations are less affected by the formation of the MW bulge compared to the cold-orbits stars~\citep[see also,][]{2018A&A...616A.180F,2019A&A...628A..11D}. The same process is predicted in the MW disc in the vicinity of the spiral arms~\citep{2018A&A...611L...2K} and confirmed by recent \textit{Gaia} data showing the most prominent MW spiral structure among younger and kinematically-colder more metal-rich stars~\citep{2022A&A...663A..38K,2022A&A...666L...4P}. Therefore, the kinematic properties of the halo tracers play a crucial role in their response to the underlying gravitation potential perturbations, including the infall of the LMC.

\section{Summary}
\label{sec:Summary}

In this paper, we have analysed the line-of-sight kinematics of 50 satellite galaxies of our Galaxy with known radial velocities.
We use the dipole and quadrupole expansions of the velocity field depending on the distance from the MW.
The dipole term describes the bulk motion, mainly determined by the motion of the Sun in the Galaxy, 
while the quadrupole term is related to the tidal interaction.

The significance of the quadrupole term of the velocity field turns out to be less than 3 sigma 
and is completely determined by the kinematics of the nearest satellites, $\DMW < 30$~kpc.
It seems natural that the kinematics of these six dwarf galaxies are most affected by tidal forces and dynamic friction in the halo of our Galaxy.

Spatial variations of the running solar apex~(dipole term) reveal the difference in the systematic motion of the nearby and far satellites.
The solar apex with respect to external satellites at distances greater than 100~kpc is $(l,b,V) = (-90 \pm 14\degr, -10.0 \pm 6.4\degr, 249\pm37~\kms{})$.
This agrees well with the known motion of the Sun in the Galaxy (the significance of the deviation is 1.7-sigma).
Therefore, we can conclude that the external satellites show an isotropic distribution of the radial velocities with a net zero motion relative to the centre of the MW and the line-of-sight dispersion of 84~\kms{} in agreement with the standard concepts of the galaxy group kinematics and the \LCDM{} model predictions.

The behaviour of the internal satellites, at distances less than 100~kpc, turns out to be more intriguing. 
The apex amplitude determined with respect to these galaxies reaches $450 \pm 54$~\kms{}, which significantly exceeds the velocity of the Sun in the Galaxy.
This corresponds to a prominent `motion' of the MW with a speed up to $226\pm50$~\kms{} relative to its satellites in direction $(l,b) = (+69 \pm 16\degr, -30 \pm 10\degr)$.

On the one hand, the almost perpendicular orientation of the velocity dipole to the direction towards the centre of the Galaxy is in favour of the rotation of the internal subsystem of satellites. This behaviour is somewhat similar to the rotation of planar structures discovered in both the MW and the Andromeda galaxies~\citep{2013MNRAS.435.2116P,2013Natur.493...62I}.
However, the extremely fast rotation would be required to explain the observed dipole.
Our estimates suggest that the circular velocity should be about $970\pm260$~\kms{}, which is highly implausible. 
Therefore, the observed behaviour of the running solar apex can not be explained by taking into account rotation of the MW satellite galaxies only.

A more realistic explanation seems to be the coherent motion of the satellites caused by the first passage of dwarf galaxies into the halo of the MW.

The behaviour of the smoothed solar apex suggests that the peculiarity of the velocity field of the nearby satellites is closely related to the LMC. The dipole term, additional to the solar apex, reaches a maximum of $226\pm50$~\kms{} with respect to the layer of satellites $35 \lesssim \DMW \lesssim 80$~kpc centred on the LMC. Carina~II and Triangulum~II make the largest contribution to this extra velocity. It is worth noting that among the 8 galaxies that have the greatest perturbing effect on the dipole, half are associated with the LMC. This is the LMC itself, and Carina~II, Carina~III and Grus~II, which are the members of the LMC group, according to \citet{2016MNRAS.461.2212J} and \citet{2018ApJ...867...19K}. This supports the hypothesis that the behaviour of the apex is associated with the first flyby of the LMC group around the MW and the perturbing effect of the massive LMC on the kinematics of other satellites.

Using two different approaches, we showed that the observed radial velocity pattern can be explained if the internal satellites were recently perturbed by a massive LMC. First, by simulating the orbital evolution of the MW satellites backwards~(with LMC) and forward~(without LMC), we 'undo' the effect of the perturbation from the massive LMC. We found that the dipole component of the velocity vanished for the inner population of the MW satellites if the LMC has a total mass of $\approx 2\times 10^{11}$~\Msun~(see Fig.~\ref{fig:apex_back_forw}) implying a genetic link of this kinematic feature with the LMC-induced perturbation of the satellites orbits.

Second, in the self-consistent high-resolution $N$-body simulations of the MW-LMC interaction, we found that halo tracers~(in our case DM particles) show the observed behaviour of the solar apex once we select particles with high angular momentum~(see Fig.~\ref{fig:Apex_AMdistr}). 
Since the MW satellites exhibit an excess of tangential velocity, and therefore a higher angular momentum, they are also more sensitive to the infall of the massive LMC.

\section*{Acknowledgements}
We thank the referee for constructive and useful comments that helped to improve this paper.
Part of the work was made using the unique scientific facility the Big Telescope Alt-azimuthal SAO RAS with the financial support of grant \textnumero~075--15--2022--262 (13.MNPMU.21.0003) of the Ministry of Science and Higher Education of the Russian Federation.

\section*{Data availability}

The observational data underlying this article are available in the article. 
The output of simulations will be made available upon reasonable request.

\bibliographystyle{mnras}
\bibliography{ref}

\newpage
\onecolumn
\appendix

\section{List of the Milky Way satellites}
\label{sec:SatelliteList}

Table~\ref{tab:SatelliteList} compiles following information.
The first column contains the common name of the galaxy.
The equatorial coordinates are given in the second and third columns in degrees.
The distance modulus, $(m-M)_0$, and corresponding distance in kpc to the galaxy are presented in the 4th an 5th columns.
The 6th column contains the heliocentric line-of-sight velocity with its errors.
Most of the velocities are taken from \citet{2020AJ....160..124M}, otherwise the reference is given in the 7th column.

{
\renewcommand{\arraystretch}{1.3}

\begin{longtable}{llrcrrll}
\caption{List of the Milky Way satellites\label{tab:SatelliteList}}\\
\hline\hline
Name              & 
\multicolumn{1}{c}{R.A.}      & 
\multicolumn{1}{c}{Dec}       & 
\multicolumn{1}{c}{$(m-M)_0$} &   
\multicolumn{1}{c}{$D$}    &
\multicolumn{2}{c}{$\Vh$}  &
Reference $^{\sharp}$  \\
&
\multicolumn{2}{c}{J2000.0}   &
\multicolumn{1}{c}{mag}   &
\multicolumn{1}{c}{kpc}  & 
\multicolumn{2}{c}{\kms{}}\\
\hline
\endfirsthead
\caption{List of the satellites (continue)}\\
\hline\hline
Name              & 
\multicolumn{1}{c}{R.A.}      & 
\multicolumn{1}{c}{Dec}       & 
\multicolumn{1}{c}{$(m-M)_0$} &   
\multicolumn{1}{c}{$D$}    &
\multicolumn{2}{c}{$\Vh$}    &
Reference $^{\sharp}$ \\
&
\multicolumn{2}{c}{J2000.0}   &
\multicolumn{1}{c}{mag}   &
\multicolumn{1}{c}{kpc}  & 
\multicolumn{2}{c}{\kms{}}\\
\hline
\endhead

\hline
\endfoot

\hline
\endlastfoot

Tucana IV              & \phantom{00}0.73    & $-60.85$\phantom{00} & 18.41           &   47.9 & $  15.9$ & $_{ -1.7}^{ +1.8}$ &                             \\ 
SMC                    & \phantom{0}13.1583  & $-72.8002$           & 18.99           &   62.8 & $ 158.0$ & $_{ -4.0}^{ +4.0}$ & \citet{2012AstBu..67..115K} \\ 
Sculptor               & \phantom{0}15.0388  & $-33.7090$           & 19.67           &   85.9 & $ 111.4$ & $_{ -0.1}^{ +0.1}$ &                             \\ 
Cetus II               & \phantom{0}19.47    & $-17.42$\phantom{00} & 17.1\phantom{0} &   26.3 &          &                    &                             \\ 
Phoenix                & \phantom{0}27.7762  & $-44.4447$           & 23.06           &  409.3 & $ -21.2$ & $_{ -1.0}^{ +1.0}$ &                             \\ 
DELVE 2                & \phantom{0}28.772   & $-68.253$\phantom{0} & 19.26           &        &          &                    &{\citet{2022A&A...657A..54B}}\\ 
Cetus III              & \phantom{0}31.3308  & $ -4.270$\phantom{0} & 22.0\phantom{0} &  251.2 &          &                    &                             \\ 
Triangulum II          & \phantom{0}33.3225  & $ 36.1784$           & 17.4\phantom{0} &   30.2 & $-381.7$ & $_{ -1.1}^{ +1.1}$ &                             \\ 
Segue 2                & \phantom{0}34.8167  & $ 20.1753$           & 17.7\phantom{0} &   34.7 & $ -40.2$ & $_{ -0.9}^{ +0.9}$ & \citet{2022ApJ...940..136P} \\ 
Eridanus III           & \phantom{0}35.6896  & $-52.2836$           & 19.7\phantom{0} &   87.1 &          &                    &                             \\ 
DES J0225+0304         & \phantom{0}36.4267  & $  3.0695$           & 16.88           &   23.8 &          &                    &                             \\ 
Hydrus I               & \phantom{0}37.3892  & $-79.3089$           & 17.2\phantom{0} &   27.5 & $  80.4$ & $_{ -0.6}^{ +0.6}$ &                             \\ 
Fornax                 & \phantom{0}39.9972  & $-34.4492$           & 20.84           &  147.2 & $  55.2$ & $_{ -0.1}^{ +0.1}$ &                             \\ 
Horologium I           & \phantom{0}43.8820  & $-54.1188$           & 19.5\phantom{0} &   79.4 & $ 112.8$ & $_{ -2.6}^{ +2.5}$ &                             \\ 
Horologium II          & \phantom{0}49.1338  & $-50.0181$           & 19.46           &   78.0 & $ 168.7$ & $_{-12.6}^{+12.9}$ &                             \\ 
Reticulum II           & \phantom{0}53.9256  & $-54.0492$           & 17.4\phantom{0} &   30.2 & $  64.7$ & $_{ -0.8}^{ +1.3}$ &                             \\ 
Eridanus II            & \phantom{0}56.0878  & $-43.5332$           & 22.9\phantom{0} &  380.2 & $  75.6$ & $_{ -1.3}^{ +1.3}$ & \citet{2022ApJ...940..136P} \\ 
Reticulum III          & \phantom{0}56.36    & $-60.45$\phantom{00} & 19.81           &   91.6 & $ 274.2$ & $_{ -7.4}^{ +7.5}$ &                             \\ 
Pictor I               & \phantom{0}70.9475  & $-50.283$\phantom{0} & 20.3\phantom{0} &  114.8 &          &                    &                             \\ 
LMC                    & \phantom{0}80.8942  & $-69.7561$           & 18.5\phantom{0} &   50.1 & $ 278.0$ & $_{ -2.0}^{ +2.0}$ & \citet{2012AstBu..67..115K} \\ 
Columba I              & \phantom{0}82.8570  & $-28.0425$           & 21.3\phantom{0} &  182.0 & $ 153.7$ & $_{ -4.8}^{ +5.0}$ &                             \\ 
Carina                 &           100.4030  & $-50.9664$           & 20.11           &  105.2 & $ 222.9$ & $_{ -0.1}^{ +0.1}$ &                             \\ 
Pictor II              &           101.18    & $-59.897$\phantom{0} & 18.3\phantom{0} &   45.7 &          &                    &                             \\ 
Carina II              &           114.1067  & $-57.9992$           & 17.79           &   36.1 & $ 477.2$ & $_{ -1.2}^{ +1.2}$ &                             \\ 
Carina III             &           114.6298  & $-57.8997$           & 17.22           &   27.8 & $ 284.6$ & $_{ -3.1}^{ +3.4}$ &                             \\ 
Ursa Major II          &           132.8741  & $ 63.1342$           & 17.5\phantom{0} &   31.6 & $-116.5$ & $_{ -1.9}^{ +1.9}$ &                             \\ 
Leo T                  &           143.7225  & $ 17.0514$           & 23.1\phantom{0} &  416.9 & $  38.1$ & $_{ -2.0}^{ +2.0}$ &                             \\ 
Antlia II              &           143.8868  & $-36.7673$           & 20.6\phantom{0} &  131.8 & $ 288.8$ & $_{ -0.4}^{ +0.4}$ & \citet{2021ApJ...921...32J} \\ 
Segue 1                &           151.7650  & $ 16.0778$           & 16.8\phantom{0} &   22.9 & $ 208.5$ & $_{ -0.9}^{ +0.9}$ &                             \\ 
Leo I                  &           152.1171  & $ 12.3064$           & 22.02           &  253.5 & $ 282.5$ & $_{ -0.1}^{ +0.1}$ &                             \\ 
Sextans dSph           &           153.2624  & $ -1.6145$           & 19.67           &   85.9 & $ 224.2$ & $_{ -0.1}^{ +0.1}$ &                             \\ 
Ursa Major I           &           158.7033  & $ 51.9350$           & 19.93           &   96.8 & $ -55.3$ & $_{ -1.4}^{ +1.4}$ &                             \\ 
Willman 1              &           162.3421  & $ 51.0519$           & 17.9\phantom{0} &   38.0 & $ -12.8$ & $_{ -1.0}^{ +1.0}$ & \citet{2022ApJ...940..136P} \\ 
Leo II                 &           168.3724  & $ 22.1540$           & 21.84           &  233.3 & $  78.0$ & $_{ -0.1}^{ +0.1}$ &                             \\ 
Leo V                  &           172.79    & $  2.22$\phantom{00} & 21.46           &  195.9 & $ 170.9$ & $_{ -1.9}^{ +2.1}$ & \citet{2021ApJ...920...92J} \\ 
Leo IV                 &           173.2367  & $ -0.5300$           & 20.94           &  154.2 & $ 132.3$ & $_{ -1.4}^{ +1.4}$ &                             \\ 
Crater, Laevens 1      &           174.0667  & $-10.8776$           & 20.8\phantom{0} &  144.5 & $ 149.3$ & $_{ -1.2}^{ +1.2}$ & \citet{2015ApJ...810...56K} \\ 
Crater II              &           177.31    & $-18.413$\phantom{0} & 20.35           &  117.5 & $  89.3$ & $_{ -0.3}^{ +0.3}$ & \citet{2021ApJ...921...32J} \\ 
Virgo I                &           180.04    & $ -0.68$\phantom{00} & 19.8\phantom{0} &   91.2 &          &                    &                             \\ 
Hydra II               &           185.4254  & $-31.9853$           & 20.64           &  134.3 & $ 303.1$ & $_{ -1.4}^{ +1.4}$ &                             \\ 
Coma Berenices, Coma I &           186.7434  & $ 23.9117$           & 18.13           &   42.3 & $  98.1$ & $_{ -0.9}^{ +0.9}$ &                             \\ 
Centaurus I            &           189.585   & $-40.902$\phantom{0} & 20.33           &  116.3 &          &                    &                             \\ 
Canes Venatici II      &           194.2896  & $ 34.3228$           & 21.02           &  160.0 & $-128.9$ & $_{ -1.2}^{ +1.2}$ &                             \\ 
Canes Venatici I       &           202.0123  & $ 33.5558$           & 21.69           &  217.8 & $  30.9$ & $_{ -0.6}^{ +0.6}$ &                             \\ 
Bootes III             &           209.3     & $ 26.8$\phantom{000} & 18.35           &   46.8 & $ 197.5$ & $_{ -3.8}^{ +3.8}$ & \citet{2012AstBu..67..115K} \\ 
Bootes II              &           209.5333  & $ 12.8483$           & 18.1\phantom{0} &   41.7 & $-117.0$ & $_{ -5.2}^{ +5.2}$ &                             \\ 
Bootes I               &           210.0209  & $ 14.5041$           & 19.11           &   66.4 & $ 101.8$ & $_{ -0.7}^{ +0.7}$ & \citet{2022ApJ...940..136P} \\ 
Ursa Minor             &           227.2854  & $ 67.2225$           & 19.4\phantom{0} &   75.9 & $-246.9$ & $_{ -0.1}^{ +0.1}$ &                             \\ 
Bootes IV              &           233.6892  & $ 43.7261$           & 21.6\phantom{0} &  208.9 &          &                    &                             \\ 
Draco II               &           238.1983  & $ 64.5653$           & 16.67           &   21.6 & $-342.5$ & $_{ -1.2}^{ +1.1}$ &                             \\ 
DELVE 1                &           247.725   & $ -0.972$\phantom{0} & 16.39           &   19.0 &          &                    & \citet{2020ApJ...890..136M} \\ 
Hercules               &           247.7650  & $ 12.79$\phantom{00} & 20.84           &  147.2 & $  46.4$ & $_{ -1.3}^{ +1.3}$ & \citet{2020MNRAS.496.1092G} \\ 
Draco                  &           260.0557  & $ 57.9271$           & 19.4\phantom{0} &   75.9 & $-291.0$ & $_{ -0.1}^{ +0.1}$ &                             \\ 
Sagittarius dSph       &           283.8312  & $-30.5453$           & 17.1\phantom{0} &   26.3 & $ 140.0$ & $_{ -2.0}^{ +2.0}$ & \citet{2012AstBu..67..115K} \\ 
Sagittarius II         &           298.1730  & $-22.0651$           & 19.32           &   73.1 & $-177.2$ & $_{ -0.6}^{ +0.5}$ & \citet{2021MNRAS.503.2754L} \\ 
Indus II $^{\dagger}$  &           309.72    & $-46.16$\phantom{00} & 21.65           &  213.8 &          &                    &                             \\
Indus I, Kim 2         &           317.2082  & $-51.1635$           & 20.0\phantom{0} &  100.0 &          &                    &                             \\ 
Grus II                &           331.02    & $-46.44$\phantom{00} & 18.62           &   53.0 & $-110.0$ & $_{ -0.5}^{ +0.5}$ &                             \\ 
Pegasus III            &           336.0942  & $  5.42$\phantom{00} & 21.56           &  205.1 & $-222.9$ & $_{ -2.6}^{ +2.6}$ &                             \\ 
Aquarius II            &           338.4813  & $ -9.3274$           & 20.16           &  107.6 & $ -71.1$ & $_{ -2.5}^{ +2.5}$ &                             \\ 
Tucana II              &           342.9796  & $-58.5689$           & 18.8\phantom{0} &   57.5 & $-129.1$ & $_{ -3.5}^{ +3.5}$ &                             \\ 
Grus I                 &           344.1765  & $-50.1633$           & 20.4\phantom{0} &  120.2 & $-140.5$ & $_{ -1.6}^{ +2.4}$ &                             \\ 
Pisces II              &           344.6292  & $  5.9525$           & 21.31           &  182.8 & $-226.5$ & $_{ -2.7}^{ +2.7}$ &                             \\ 
Tucana V               &           354.35    & $-63.27$\phantom{00} & 18.71           &   55.2 & $ -36.2$ & $_{ -2.2}^{ +2.5}$ &                             \\ 
Phoenix II             &           354.9975  & $-54.406$\phantom{0} & 19.6\phantom{0} &   83.2 & $  32.4$ & $_{ -3.8}^{ +3.7}$ &                             \\ 
Tucana III             &           359.15    & $-59.6$\phantom{000} & 17.01           &   25.2 & $-102.3$ & $_{ -0.4}^{ +0.4}$ & \citet{2022ApJ...940..136P} \\ 

\hline
\multicolumn{8}{l}{$^{\sharp}$ most of the data were taken from \citet{2020AJ....160..124M}} \\
\multicolumn{8}{l}{$^{\dagger}$ probably not a galaxy \citep{2021ApJ...916...81C}} \\

\hline\hline
\end{longtable}

}

% Don't change these lines
\bsp	% typesetting comment
\label{lastpage}
\end{document}